\documentclass[fleqn,usenatbib]{mnras}

\usepackage{newtxtext,newtxmath}

\usepackage[T1]{fontenc}

\DeclareRobustCommand{\VAN}[3]{#2}
\let\VANthebibliography\thebibliography
\def\thebibliography{\DeclareRobustCommand{\VAN}[3]{##3}\VANthebibliography}

\usepackage{float}

\newcommand{\chone}{$3.6\mu{\rm m}$}
\newcommand{\chtwo}{$4.5\mu{\rm m}$}
\newcommand{\Zsun}{${\rm Z}_{\odot}$}
\newcommand{\cii}{{\sc [CII]}}

\newcommand{\oiiio}{{\sc [OIII]} $\lambda \lambda 4959,5007$~}
\newcommand{\hboii}{{\sc H}$\beta$ + {\sc [OIII]}}
\newcommand{\halpha}{{\sc H}$\alpha$}

\newcommand{\av}{${A_{\rm V}}$}
\newcommand{\ew}{${\rm EW}_{\rm 0}$}
\newcommand{\irxb}{${\rm IRX}$--$\beta$}
\newcommand{\irx}{${\rm IRX}$}
\newcommand{\muv}{${M_{\rm UV}}$}
\newcommand{\Lsun}{${\rm L}_{\odot}$}
\newcommand{\Msun}{${\rm M}_{\odot}$}
\newcommand{\Lir}{$L_{\rm IR}$}
\newcommand{\Luv}{$L_{\rm UV}$}
\newcommand{\sfrunit}{${\rm M}_{\odot}/{\rm yr}$}
\newcommand{\mstar}{$M_{\star}$}
\newcommand{\lmstar}{${\rm log}_{10}(M_{\star}/{\rm M}_{\odot})$}

\usepackage{graphicx}	
\usepackage{amsmath}	
\usepackage{subcaption}

\title[Spatially resolved $z \simeq 5$ CRISTAL galaxies with PRIMER]{JWST PRIMER: A lack of outshining in four normal $\mathbf{z =4}-\mathbf{6}$ galaxies from the ALMA-CRISTAL Survey}

\author[N. E. P. Lines]{N. E. P. Lines,$^{1,2}$\thanks{E-mail: natalie.lines@port.ac.uk}
R. A. A. Bowler,$^{1}$\thanks{E-mail: rebecca.bowler@manchester.ac.uk}
N. J. Adams,$^{1}$
R. Fisher,$^{1}$
R. G. Varadaraj,$^{3}$
Y. Nakazato,$^{4}$\newauthor
M. Aravena,$^{5}$
R. J. Assef,$^{5}$
J. E. Birkin,$^{6}$
D. Ceverino,$^{7,8}$
E. da Cunha,$^{9,10}$
F. Cullen,$^{12}$
I. De Looze,$^{11}$\newauthor
C. T. Donnan,$^{12}$
J. S. Dunlop,$^{12}$
A. Ferrara,$^{13}$
N. A. Grogin,$^{14}$
R. Herrera-Camus,$^{15}$
R. Ikeda,$^{16,17}$\newauthor
A. M. Koekemoer,$^{14}$
M. Killi,$^{5}$
J. Li,$^{9,10,18}$
D. J. McLeod,$^{12}$
R. J. McLure,$^{12}$
I. Mitsuhashi,$^{17,19}$ \newauthor
P. G. P\'erez-Gonz\'alez,$^{20}$ 
M. Relano,$^{21}$
M. Solimano,$^{5}$
J.~S.~Spilker,$^{6}$
V. Villanueva,$^{15}$
N. Yoshida$^{22,23}$\\
\vspace{1cm}
{\it \normalsize Affiliations are listed at the end of the paper}
}

\date{Accepted XXX. Received YYY; in original form ZZZ}

\pubyear{2024}

\begin{document}
\label{firstpage}
\pagerange{\pageref{firstpage}--\pageref{lastpage}}

\maketitle
\begin{abstract}
We present a spatially resolved analysis of four star-forming galaxies at $z = 4.44-5.64$ using data from the~\emph{JWST} PRIMER and ALMA-CRISTAL surveys to probe the stellar and inter-stellar medium properties on the sub-${\rm kpc}$ scale.
In the $1-5\,\mu{\rm m}$~\emph{JWST} NIRCam imaging we find that the galaxies are composed of multiple clumps (between $2$ and $\sim 8$) separated by $\simeq 5\,{\rm kpc}$, with comparable morphologies and sizes in the rest-frame UV and optical.
Using {\tt BAGPIPES} to perform pixel-by-pixel SED fitting to the~\emph{JWST} data, we show that the SFR ($\simeq 25\,{\rm M}_{\odot}/{\rm yr}$) and stellar mass (\lmstar \space $\simeq 9.5$) derived from the resolved analysis are in close ($ \lesssim 0.3\,{\rm dex}$) agreement with those obtained by fitting the integrated photometry.
In contrast to studies of lower-mass sources, we thus find a reduced impact of outshining of the older (more massive) stellar populations in these normal $z \simeq 5$ galaxies.
Our~\emph{JWST} analysis recovers bluer rest-frame UV slopes ($\beta \simeq -2.1$) and younger ages ($\simeq 100\,{\rm Myr}$) than archival values.
We find that the dust continuum from ALMA-CRISTAL seen in two of these galaxies correlates, as expected, with regions of redder rest-frame UV slopes and the SED-derived \av, as well as the peak in the stellar mass map.
We compute the resolved \irxb~relation, showing that the IRX is consistent with the local starburst attenuation curve and further demonstrating the presence of an inhomogeneous dust distribution within the galaxies.
A comparison of the CRISTAL sources to those from the FirstLight zoom-in simulation of galaxies with the same \mstar~and SFR reveals similar age and colour gradients, suggesting that major mergers may be important in the formation of clumpy galaxies at this epoch.  
\end{abstract}




\section{Introduction}

A full understanding of the formation and evolution of early galaxies requires spatially resolved information.
An analysis of the emitted rest-frame UV and optical light provides a measure of the stellar properties at the time of observation, as well as the star formation history (SFH) of the galaxy by probing stars of different ages.
The continuum and line emission in this wavelength range can also be impacted by (potentially inhomogeneous) dust attenuation (e.g.~\citealp{AbdurroufCoe23, Hashimoto23, Killi24}) and/or metallicity gradients (e.g. identified through spectroscopic studies;~\citealp{Tripodi24, Venturi24}). 
The wavelength dependent morphology further provides information on the formation mechanism of high-redshift galaxies by highlighting any signatures of merging activity, for example through the identification of multiple discrete clumps~\citep{Bowler17, Barisic17, Hashimoto19}, or established stellar disks through the detection of a centralised redder component attributed to older stars~\citep{Ji23, Setton24, Fujimoto24}.

The~\emph{Hubble Space Telescope (HST)} has provided a wealth of information about the morphology and colour of galaxies up to very high redshifts ($z \lesssim 11$, e.g.~\citealp{Oesch10, Holwerda15}). 
However, at $z > 5$ the~\emph{HST}/Wide-Field Camera 3 (WFC3) imaging is limited to probing the rest-frame UV regime, where dust obscuration can play a significant role.
It is therefore not clear if the clumpy morphology observed in many $z \gtrsim 5$ galaxies is a result of mergers or if these components are embedded in an underlying disk (e.g.~\citealp{Barisic17}).
With the advent of imaging from the~\emph{James Webb Space Telescope (JWST)} NIRCam instrument, it is possible for the first time to obtain sub-kpc scale imaging in the rest-frame optical for galaxies up to $z \simeq 10$.
These measurements provide a step change in the study of the rest-frame optical continuum and emission line properties at these redshifts, which were previously only inferred from relatively poor resolution~\emph{Spitzer}/IRAC photometry (e.g.~\citealp{Stefanon23}).

The majority of known $z > 4$ galaxies have been identified via the detection of the Lyman-$\alpha$ break in the rest-frame UV, sometimes in combination with a narrow-band filter to pick up the Lyman-$\alpha$ emission line itself (see~\citealp{Stark16} for a review).
The physical properties of the galaxies are then obtained through spectral energy distribution (SED) fitting of the observed optical and near-infrared (NIR) photometry.
Due to the compact sizes of the majority of galaxies at $z \gtrsim 4$, even in~\emph{JWST} imaging the galaxies are marginally resolved (with half-light radii of $\sim 1\,{\rm kpc}$;~\citealp{Morishita24, Ono24, Varadaraj24}).
Global properties such as the stellar mass and star-formation rate (SFR) are therefore derived from the integrated light of the source.
While the biases that can occur in properties derived from SED fitting have been extensively discussed for galaxies at $ z \lesssim 2$ (e.g~\citealp{Sorba18, Leja19}), it is not clear how discrepant resolved and integrated properties of galaxies might be at high redshift.
For example, the reduced cosmic time available and the potential lack of dust attenuation~\citep{Fudamoto20a, Bowler24} could in principle simplify the SED fitting process making globally derived properties robust.
With the unique combination of sensitivity, spatial resolution and spectroscopic capabilities of~\emph{JWST} at $>1\,\mu{\rm m}$, it is now possible to perform spatially resolved analyses that probe the rest-frame optical for the first time. This allows the testing of the previously obtained physical properties for a wide range of $z \gtrsim 4$ galaxies.

Using~\emph{JWST} NIRCam imaging of five lensed $z \simeq 5.2$--$8.5$ galaxies,~\citet{GimenezArteaga23} found that when spatially resolving the galaxies, the rest-frame UV light was dominated by compact regions that show young ages, as revealed by strong \oiiio emission lines.
The resolved SED fitting revealed regions of the galaxies with substantial stellar mass, but less current star formation and older ages.
While the light from these regions is measured in the integrated photometry, these areas of high stellar mass are `outshone' by the younger pockets of star formation, which then leads to an underestimate of the stellar mass when fitting to the integrated photometry by up to $1\,{\rm dex}$.
A similar effect is seen in the strongly lensed galaxy nicknamed the `Cosmic Grapes' introduced in~\citet{Fujimoto24}. 
In this galaxy, ~\citet{GimenezArteaga24} demonstrate the effect of outshining and the inability of a variety of typically assumed SFHs in reproducing the galaxy properties from the global photometry.
These studies, based on~\emph{JWST}-selected strongly lensed galaxies, demonstrate that outshining appears to be significant at \lmstar $< 9$ in `typical' sources (lying on the galaxy main sequence).
It might be expected that such effects would be reduced at higher stellar masses, where in general the galaxy is experiencing less extreme or dominant pockets of very young star formation.
Indeed,~\citet{PerezGonzalez23} find no systematic offset between stellar masses derived from a resolved and integrated SED fitting analysis using a sample of red galaxies at $z > 3$ with \lmstar$\simeq 10$.
Further results at $z \simeq 4$ indicate a reduced effect of outshining at \lmstar $\simeq 10.5$, recovering similar spatially resolved properties to those obtained from fitting the integrated/global photometry~\citep{Nelson23, Setton24}.
The goal of this work is to connect these studies, which probe widely different mass ranges and are based on various sample selections, by determining the significance of outshining in `typical' main-sequence galaxies at $z \simeq 5$.

The ALMA-[CII] resolved ISM in star-forming galaxies with ALMA (CRISTAL: Herrera-Camus in prep.) large program aims to revolutionise studies of the rest-frame UV to FIR SEDs of $z \simeq 5$ galaxies, with high-resolution ($\sim 0.3$ arcsec, equivalent to $\sim 2.0$ kpc at $z=4.5$) \cii~and dust continuum observations in a sample of over 20 main-sequence galaxies.
Previous observations of the FIR emission in high-redshift galaxies has shown that the stellar mass range of \lmstar $= 9$--$10$ is key for understanding the build-up of dust at early times.
As shown in samples of galaxies observed through various ALMA large programs, within this mass range the obscured SFR fraction of $z = 4$--$8$ galaxies is $\simeq 0.5$~\citep{Fudamoto20, Algera23, Mitsuhashi24, Bowler24}, with some sources showing $f_{\rm obs} > 0.9$~\citep{Fudamoto21}.

CRISTAL has already revealed significant dust obscured star formation~\citep{Mitsuhashi23} in typical $z =4$--$6$ galaxies.
It has also discovered the complex highly-resolved \cii~emission around two massive galaxies suggestive of a merging event (\citealt{Posses24}; \citealt{Solimano24}), along with the first significant sample of resolved rest-frame UV to FIR SED fitting in Li et al. (2024), dust temperature measurements in HZ10~\citep{Villanueva24} and the measurement of \cii~sizes~\citep{Ikeda24}.

In this work we perform a spatially resolved SED fitting analysis of four normal $z \simeq 4$--$6$ galaxies with \lmstar\space$ \simeq 9.5$ to investigate the effect of outshining in main-sequence star-forming galaxies and to understand the trends and potential biases in the previously obtained SED fitting properties of galaxies of this stellar mass.
Our sample includes all of the sources that are part of the CRISTAL ALMA large program that overlap with the Public Release Imaging for Extragalactic Research (PRIMER;~\citealp{Dunlop21})~\emph{JWST} imaging, which observed the Cosmic Evolution Survey (COSMOS) field with eight NIRCam filters.
These sources are extended on the scale of an arcsecond and are detected with a high signal-to-noise ratio (SNR), allowing a detailed spatially resolved analysis.
The resolved parameters are then compared with high-spatial resolution ($\sim 0.25$ arcsec) ALMA-CRISTAL dust continuum emission.

The paper is structured as follows.
We present the datasets we use in Section~\ref{sect:data}, followed by a description of our point spread function (PSF) homogenisation and resolved SED fitting analysis in Section~\ref{sect:method}.
The spatially resolved SED fitting maps are presented in Section~\ref{sect:results}, along with a comparison of the resolved and integrated properties.
In Section~\ref{sect:discussion} we discuss our results and compute a resolved infrared-excess ({\irx} $= \log_{10}[$\Lir/\Luv$]$)-$\beta$ relation for the sample (where $\beta$ is the rest-frame UV slope; $F_{\lambda} \propto \lambda^{\beta}$).
We end with our conclusions in Section~\ref{sect:conclusions}.
We assume AB magnitudes throughout, and use the cosmological parameters $\Omega_M=0.3,~\Omega_{\Lambda}=0.7$ and $H_0=70{\rm~km~s^{-1}~Mpc^{-1}}$.
At $z = 4.5~(5.5)$, 1 arcsec corresponds to a physical distance of $6.6~(6.0)\,{\rm kpc}$.

\section{Data and Sample}\label{sect:data}
This study is based on~\emph{JWST} imaging data from the PRIMER survey, which allows a pixel-by-pixel spatially resolved analysis of the four galaxies we consider.
We then compare the resulting physical parameter maps to ALMA observations in Band 7 obtained as part of the CRISTAL large program.
In this section we describe the main features of these datasets and the selection function of the four galaxies we study.

\subsection{\emph{JWST} PRIMER}
PRIMER was a Cycle 1 program (PI: Dunlop, PID 1837) providing deep NIRCam imaging within the COSMOS and Ultra-Deep Survey (UDS) fields.
We utilise the PRIMER internal V0.5 data release, which includes the first half of the dataset in the PRIMER-COSMOS field as observed in December 2022 and January 2023 (known as the COSMOS-2 observations).
Further data, observed in April and May 2023 covered the second half, or COSMOS-1 observations, however the four sources in our analysis sit within the COSMOS-2 region and hence this additional data was not required for this analysis.
The data was reduced using a custom version of the~\emph{JWST} calibration pipeline, PENCIL (PRIMER enhanced NIRCam Image Processing Library; Magee in prep.), starting from the {\sc uncal.fits} files.
The reduction process removed excess $1/f$ striping, masked the residual snowball artefacts and removed wisp artefacts in the F150W and F200W filters.
The astrometry was aligned to GAIA data release 3~\citep{GaiaCollaborationVallenari23}. 
The COSMOS-2 data we use consists of a mosaic of 19 NIRCam pointings across 8-band photometry (F090W, F115W, F150W, F200W, F277W, F356W, F410M, F444W) where the $5\sigma$ limiting depths are approximately $m_{\rm AB} = 28$ (0.3 arcsec diameter aperture;~\citealp{Donnan24}).
The mosaic has a pixel scale of 0.03 arcsec per pixel.

We verified the internal astrometric accuracy of the data using cut-outs of the size of $3000\times3000$ pixels (or $1.5\times1.5$ arcmin) centred on our targets. 
Using the sample of sources selected within these cut-outs, we determined that the relative image alignment has a precision of $\lesssim 0.02$ arcsec or less than one pixel.
For our analysis we homogenised the PSF of the images using the {\tt WebbPSF} models \citep{Perrin12,Perrin14}.
Each filter image was matched to that of the F444W PSF using convolution kernels derived from the PSFs using the Wiener-Hunt deconvolution algorithm. 

\subsection{ALMA-CRISTAL large program}
The four galaxies that we study in this work were observed as part of the ALMA-CRISTAL large program (2021.1.00280.L; PI: Rodrigo Herrera-Camus, see Herrera-Camus et al. in prep) in Cycle 8.
The goal of CRISTAL was to provide high sensitivity and spatial resolution observations of the \cii~line and dust continuum emission in a subset of galaxies from the ALMA Large Program to Investigate C+ at Early Times (ALPINE; 2017.1.00428.L; PI: O. Le Fèvre;~\citealp{LeFevre20, Faisst20, Bethermin20}).
Further constraints were placed on the sample selection so that the 19 CRISTAL galaxies had stellar masses derived by SED fitting using {\sc Le Phare} \citep{Arnouts99,Ilbert06} of \lmstar$>9.5$, had~\emph{HST} data available and were within a factor of three of the evolving star-formation main sequence (see Herrera-Camus et al. in prep for details).
The ALMA Band 7 observations are centred on the known frequency of the \cii $158\,\mu{\rm m}$ line as detected in ALPINE.
Extended and compact configurations were used to provide a beam size of $\sim 0.3\,{\rm arcsec}$ in the natural weighting, but with a maximum recoverable scale of $> 4\,{\rm arcsec}$.
Two of the galaxies we study in this work, CRISTAL-11 and CRISTAL-13, are detected significantly ($>3 \sigma$) in the dust continuum first by ALPINE~\citep{Bethermin20}, and now by CRISTAL.
CRISTAL-15 and CRISTAL-17 had negligible detection in the dust continuum, however we utilise upper limits on the \Lir~in our IRX analysis.
The extraction of the flux and inferred luminosity are described in~\citet{Mitsuhashi23}, and we utilise these values in our determination of the IRX, though we do not include the ALMA data in our SED fitting and therefore do not PSF match the NIRCam images to the ALMA resolution.
For CRISTAL-11 the beam of the natural weighted data was $0.50 \times 0.39$ arcsec, with a root-mean square (RMS) of $19 \mu{\rm Jy}/{\rm beam}$.
For CRISTAL-13 the beam of the natural weighted data was $0.54 \times 0.45$ arcsec, with a RMS of $15 \mu{\rm Jy}/{\rm beam}$. 
For CRISTAL-15 the beam of the natural weighted data was $0.42 \times 0.36$ arcsec, with a RMS of $11.5 \mu{\rm Jy}$.
For CRISTAL-17 the beam of the natural weighted data was $0.76 \times 0.64$ arcsec, with a RMS of $9 \mu{\rm Jy}$.
The astrometric accuracy of the dust continuum images for CRISTAL-11 and CRISTAL-13, approximated using $\simeq {\rm FWHM}/{\rm SNR}$, are $120\,{\rm milliarcsec}$ and $80\,{\rm milliarcsec}$, respectively.

\begin{table}
    \centering
      \caption{The basic information and measured total or integrated photometry for the four CRISTAL galaxies studied in this work.
      The rows show the spectroscopic redshift from the ALPINE program~\citep{LeFevre20} determined from \cii, the R.A. and Dec. of the rest-frame UV component, and the rest-frame equivalent width of Lyman-$\alpha$ as presented in the ALPINE catalogue~\citep{Faisst20, Cassata20}.
      The remaining rows show the total flux in the PRIMER~\emph{JWST} NIRCam filters measured in nJy, calculated by summing the pixels of the source across our mask and accounting for the missing flux outside this aperture.
      The errors on the photometry were determined from the standard error on the mean, however for our SED fitting analysis we set a floor on the error of 5 percent.
      } 
\label{tab:flux} 
\resizebox{\columnwidth}{!}{%
\begin{tabular}{lcccc}
\hline
ID                       & CRISTAL-11      & CRISTAL-13      & CRISTAL-15      & CRISTAL-17      \\ \hline
$z_{\rm [CII]}$          & 4.44            & 4.58            & 4.58            & 5.64            \\ \hline
R.A. [deg]                    & 150.1359        & 150.1715        & 150.1986        & 150.1630        \\
Dec. [deg]                    & 2.2579          & 2.2873          & 2.3006          & 2.4257          \\ \hline
${\rm EW}_{{\rm Ly}\alpha,\rm 0}$ [\AA]       & $34 \pm 17$     & $9.9\pm 1.2$    & $22.4 \pm 1.4$  & $58 \pm 13$     \\ \hline
F090W [nJy]& 419$\pm$4 & 342$\pm$3 & 588$\pm$2 & 166$\pm$1 \\
F115W [nJy]& 420$\pm$3 & 394$\pm$3 & 591$\pm$2 & 220$\pm$1 \\
F150W [nJy]& 411$\pm$3 & 377$\pm$2 & 548$\pm$2 & 163$\pm$1 \\
F200W [nJy]& 563$\pm$2 & 548$\pm$2 & 623$\pm$2 & 141$\pm$1 \\
F277W [nJy]& 783$\pm$2 & 816$\pm$2 & 968$\pm$2 & 172$\pm$1 \\
F356W [nJy]& 811$\pm$2 & 784$\pm$2 & 733$\pm$2 & 244$\pm$1 \\
F410M [nJy]& 653$\pm$5 & 616$\pm$4 & 528$\pm$4 & 124$\pm$2 \\
F444W [nJy]& 672$\pm$5 & 623$\pm$4 & 555$\pm$4 & 196$\pm$2 \\ \hline
\end{tabular}}
\end{table}

\begin{figure*}
    \centering
    \includegraphics[width=\textwidth, trim = 4cm 3cm 4cm 1cm]{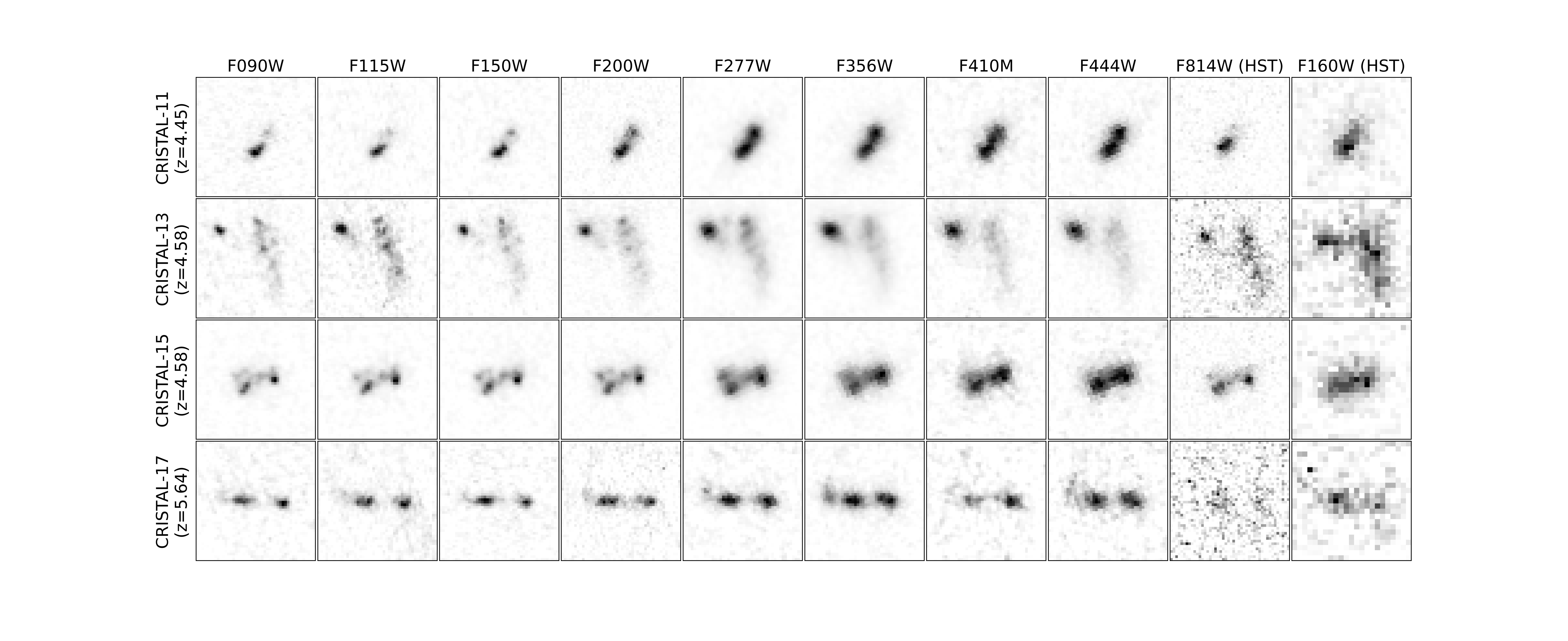}
    \caption{Postage-stamp cut-out images of the NIRCam data available for the four CRISTAL galaxies in this work.  
    The galaxies were observed as part of the CRISTAL ALMA large program, and overlap with the deep~\emph{JWST} data in the PRIMER-COSMOS field. 
    Each image is 1.4 arcsec on a side, with North to the top and East to the left.
    The stamps are ordered with wavelength from blue (left) to red (right) for the~\emph{JWST} images, followed by the previous best~\emph{HST} images in the ACS/F814W and WFC3/F160W filter bands. 
    The images displayed are from the original PRIMER reductions and are before PSF homogenisation has been conducted. 
    The pixel scale is 0.03 arcsec for all filter bands except WFC3/F160W for which it is 0.06 arcsec, and the stamps have been scaled from [-3, max]$\sigma$, where $\sigma$ was determined from the RMS of the empty background.}
    \label{fig:flux}
\end{figure*}

\subsection{Sample}\label{sect:sample}
In this study we focus on a sample of spectroscopically confirmed, massive Lyman-break galaxies at $z \simeq 4$--$6$ that are both in the CRISTAL sample and have deep multi-band NIRCam imaging from the PRIMER survey.
The sources were initially part of a large $i$-band selected sample in COSMOS, of which 118 were observed with ALMA as part of the ALPINE large program~\citep{LeFevre20}.
They were originally spectroscopically confirmed via the detection of the Lyman-$\alpha$ line\footnote{The original IDs as part of this program, and carried into the ALMA ALPINE follow-up are DEIMOS\_COSMOS\_630594 (C11), vuds\_cosmos\_5100994794 (C13), vuds\_cosmos\_5101244930 (C15), DEIMOS\_COSMOS\_742174 (C17)}, and subsequently detected in the \cii~cooling line in ALPINE to provide a systemic redshift (see Table~\ref{tab:flux}).
Several CRISTAL galaxies have~\emph{JWST} imaging or spectroscopic follow-up, however only four benefit from the full eight NIRCam filters from PRIMER.
Postage stamp images of these galaxies are shown in Fig.~\ref{fig:flux}.

As detailed in~\citet{LeFevre20} and~\citet{Faisst20}, the ALPINE sample includes galaxies at $z > 4$ selected using different techniques, with spectroscopic follow-up including sources with both rest-frame UV emission and absorption features.
Using the \cii~detections, the galaxies were given morpho-kinematic classifications as an indication of the potential formation mechanism, based on the galaxy morphology and observed kinematic structure.
We present the rest-frame equivalent-widths (\ew) from the ALPINE catalogue in Table~\ref{tab:flux}.
\citet{Cassata20} note that the Lyman-$\alpha$ emission in CRISTAL-11 and CRISTAL-15 may be offset from the optical and \cii~peak by $\sim 0.5$ arcsec (which is approximately $3$ kpc at this redshift).

CRISTAL-11 and CRISTAL-13 were initially identified using the Lyman-break technique.
They both have an assigned morpho-kinematic class of `extended dispersion-dominated', which corresponds to a resolved/extended \cii~emission line that did not show a velocity gradient in the ALPINE \cii~data.
CRISTAL-13 also has a classification of `UNC' in~\citet{Jones21}, indicating that the merger or rotating disk classications cannot be robustly assigned.
In contrast, CRISTAL-15 and CRISTAL-17 were both initially selected as narrow-band detected sources, meaning that they had significant Lyman-$\alpha$ emission to cause an excess in the narrow-band images (SuprimeCam NB711 at $z = 4.5$ and NB814 at $z = 5.7$ for the two sources respectively).
The morpho-kinematic class of CRISTAL-15 is a `pair-merger', indicating that the source shows evidence for an interaction, typically showing a disturbed velocity map or separate components in the \cii~or auxiliary data.
As stated in~\citet{Cassata20}, this source has multiple components in the optical data and hence was classified as a merger.
Finally, CRISTAL-17 has not been allocated a morpho-kinematic class from these previous works due to a weak \cii~emission line, however~\citet{Cassata20} identify that the \cii~emission is offset by 0.7 arcsec from the peak of the Lyman-$\alpha$ and optical emission.

As presented in Herrera-Camus et al. 2025, in the ALMA-CRISTAL data all four galaxies have \cii~detections with a  measured signal-to-noise of 3.7--12.
For CRISTAL-11 and -13, the \cii~peaks at the position of the dust continuum emission, but there is additional \cii~flux extending beyond the rest-frame UV/optical emission.
CRISTAL-15 has \cii~emission centered on the source, with the weak emission in CRISTAL-17 being offset to the North.
From kinematic analysis presented in Lee et al. in prep, CRISTAL-11 and CRISTAL-15 show evidence for being a rotating disk.

\section{Methods}\label{sect:method}
Here we present the key steps in our analysis of the high spatial resolution~\emph{JWST} NIRCam images from PRIMER.
Specifically, we carry out a pixel-by-pixel SED fitting analysis and estimate the rest-frame UV and optical sizes from a parametric and non-parametric fitting approach, which we then can compare to the dust continuum sizes obtained in~\citet{Mitsuhashi23}.

\subsection{Pixel fitting criteria}\label{ssec:criteria}
We created an image mask for each source that identified pixels in which we are confident that we can determine robust physical parameters from our SED fitting analysis.
The mask was created using an inverse-variance weighted (IVW) stack, combining all the available bands. 
Pixels were selected if they reached a SNR of 20 in this band-combined stack.
This criteria results in 100--450 individual pixels being fit per galaxy, with the major features retained for each source across the wide wavelength range covered.
The fitting of these pixels results in derived physical parameters that we call the resolved quantities following the terminology of~\citet{GimenezArteaga23}.
Integrated photometry was obtained for each galaxy by summing the fluxes of each individual pixel used in the resolved fits to produce a single flux value for each photometric band. 
This was done (as opposed to aperture photometry) to allow for a direct comparison of the inferred parameters in a resolved versus integrated analysis using the same pixels.
Both the resolved and integrated fits miss a proportion of the total galaxy light due to the conservative mask we use, which corresponds to a recovery fraction of 0.72, 0.69, 0.78 and 0.53 of the total light for CRISTAL-11, -13, -15 and -17 respectively.
The total flux here was determined in each band by summing the flux in the cutout image at $> 2\sigma$ significance, taking an average across the NIRCam bands available (no significant trends with wavelength were seen, as supported by the comparable rest-frame UV and optical sizes we determine).
This method for determining the total flux gives consistent results to using a fixed large aperture.
We correct both the resolved parameters and the integrated photometry by these factors.
The enclosed flux of the mask used for CRISTAL-17 contains a significantly lower fraction of the total galaxy light than the other three sources: CRISTAL-17 is the faintest galaxy in our sample and therefore a lower proportion of the pixels pass our conservative masking criterion.
As we show in Section \ref{sect:phot}, CRISTAL-17 also shows a significantly lower flux in the NIRCam data compared to the previous ground-based and~\emph{Spitzer}/IRAC photometry.
We identify a lower redshift companion galaxy to the North of CRISTAL-17 that appears to have contaminated the measured flux for the central source in the previous COSMOS2015 and COSMOS2020 catalogues (most strikingly by a factor of $> 2$ in the \chone~band), leading to the observed lower fluxes in the high resolution NIRCam data.

\subsection{SED fitting analysis} \label{ssec:setup}
We perform SED fitting to the pixel-by-pixel/resolved and integrated (summing the pixel photometry) fluxes using the eight NIRCam bands.
We do not include the ALMA data in our fitting, due to the reduced resolution of this data, however a JWST+ALMA matched resolution SED fitting analysis of the full CRISTAL sample is presented in Li et al. (2024) and we present a comparison to the results of this study in Appendix~\ref{appendix:li}.
We use {\tt BAGPIPES} (Bayesian Analysis of Galaxies for Physical Inference and Parameter EStimation;~\citealp{Carnall18}) to perform our SED fitting.
For our fiducial physical property estimates we assume a constant star formation history (CSFH) model, which allows a direct comparison with the works of~\cite{Carnall23} and~\cite{GimenezArteaga23} who follow a similar approach but for fainter galaxies.
Note that we expect the CSFH to be a good approximation to the pixel fitting as these regions likely have a simple SFH that can be approximated with different magnitude and duration top-hat functions.
However, we expect the CSFH assumption to affect the integrated photometry results to a greater degree (see Section~\ref{sect:discussion} and~\citealp{GimenezArteaga24}).
We also additionally fit with a delayed exponential model (delayed-$\tau$; $SFR \propto t\,e^{-t/\tau}$) to provide a comparison.
As shown in~\citet{GimenezArteaga24}, a similar peaked parameterisation (the double-power law) provides the greatest agreement between the integrated and resolved measurements, allowing us to test the effect of varying the SFH on our results.

We allow the age to vary from 1 Myr to 1 Gyr, and we fix the metallicity at 0.5\space \Zsun. 
The dust attenuation law is assumed to be~\citet{Calzetti00} and we allow the magnitude of attenuation to vary in the range 0 < \av < 2. 
We include nebular emission in our model using the {\tt BAGPIPES} default parametrisation and allowing the ionisation parameter $U$ to vary in the range $-4 < {\rm log}_{10}(U) < -2$. 
Nebular and stellar light is assumed to be subject to the same attenuation.
We set a floor to the error of 5 percent as is standard in SED fitting analyses.
We fix the redshifts of the four objects to the previously determined systemic spectroscopic redshifts from the ALPINE program~\citep{LeFevre20}.  
Applying a resolved methodology like that performed here allows us to observe regions of these galaxies that would otherwise be blended and fitted as a single source (with a single value of e.g. \av, age) as we discuss in detail below.
We homogenised the PSF of the images to match that of coarsest spatial resolution data, which provided a smoothing of the derived results over the scale of the F444W PSF size, which is $\sim 0.16$ arcsec or five pixels.
Therefore neighboring pixels are not independent in the fitting analysis on this scale.

The parameters inferred for each pixel fit allow us to build maps of each parameter matching the spatial resolution of the F444W image, where we display the 50{$^{\rm{th}}$} percentile from the posterior distribution derived in {\tt BAGPIPES}.
In addition to parameter estimates produced from the SED fitting, we also inferred other parameters from the best-fitting SED model in the integrated and resolved cases. 
By fitting a power-law slope in the wavelength range of $\lambda_{\rm rest} = 1268$--$2580$\AA~\citep
{Calzetti94} we extracted an estimate of the rest-frame UV slope ($\beta$, where $F_{\lambda} \propto \lambda^{\beta}$).
Rest-frame equivalent widths of H$\alpha$ and \hboii~were also derived from the best-fitting SED  using the indices fitting function in {\tt BAGPIPES}. 
We also present the empirical colour maps across the rest-frame UV and Balmer break region as obtained by subtracting the relevant broad-band images. 
For CRISTAL-11, CRISTAL-13 and CRISTAL-15 at $z \simeq 4.5$ the F090W$-$F150W colour is a proxy for the rest-UV slope and the Balmer break is bracketed by F150W$-$F277W, while for CRISTAL-17 at $z = 5.64$ the appropriate colours are F090W$-$F200W and F200W$-$F356W.
We note here that the rest-frame optical emission lines of {\sc H}$\alpha$ lies within the F356W band for CRISTAL-11, -13 and -15 and in F444W (not F410M) for CRISTAL-17.
The \hboii~complex lies within F277W for the three $ z \simeq 4.5$ sources, and in F356W for CRISTAL-17 at $z = 5.6$.

\subsection{Estimates of global properties }
We estimate the global properties of the four galaxies following the two methodologies included in~\citet{GimenezArteaga23}.
First, we compute the resolved physical properties for the sample using the results of the pixel-by-pixel fitting, which takes into account the potential differing properties across the source.
To derive the resolved stellar mass and SFR, we simply summed all pixels present in our maps that were included in the fitting analysis according to our IVW mask.
We then correct this to a measure of the total mass using the aperture corrections we derive in Section~\ref{ssec:criteria}.
The uncertainties were derived from summing in quadrature the error on the pixel values.
For the age, \av~and rest-frame UV slope, we computed the mean and the standard error from the pixel map without any weighting.
Second, we compute an integrated determination of the physical properties by summing the pixel-by-pixel fluxes to produce unresolved photometry, which was then fit with {\tt BAGPIPES}.
Here the errors are determined from the 16th and 84th percentile of the posterior distributions.
The integrated method is equivalent to aperture or model fitting photometry on the full source, and thus provides a key comparison to the resolved results to determine whether the physical properties could be biased by effects such as outshining.

\subsection{Size measurements}
We measured the rest-frame UV and optical sizes of the four CRISTAL galaxies in our sample using both a parametric and non-parametric approach.
The full details of the fitting procedure are presented in~\citet{Varadaraj24}.
Single S\'{e}rsic profiles were fit to the NIRCam images using {\sc PyAutoGalaxy}~\footnote{https://github.com/Jammy2211/PyAutoGalaxy} \citep{Nightingale23}.
We also computed non-parametric size measurements by taking the circularised radius of the brightest 50 percent of pixels above a threshold of $5 \sigma$ in the relevant filter image.
The non-parametric size estimate was corrected for the effect of the PSF by subtracting the non-parametric size determined from the PSF in quadrature.
Interestingly, all four of the galaxies in this work were classed as poor parametric fits following the analysis of~\citet{Varadaraj24}, due to their multi-component morphology cannot be adequately described with a single S\'{e}rsic profile.
We therefore present two sizes, one probing the rest-frame UV and one the rest-frame optical light, computed using the non-parametric size approach in this work.
The size of the dust continuum emission, measured in the ALMA-CRISTAL program for CRISTAL-11 and CRISTAL-13, was taken from~\citet{Mitsuhashi23} who fit a single 2D exponential profile ($n = 1$) in the visibility data.

\begin{figure*}

    \includegraphics[width=0.9\textwidth]{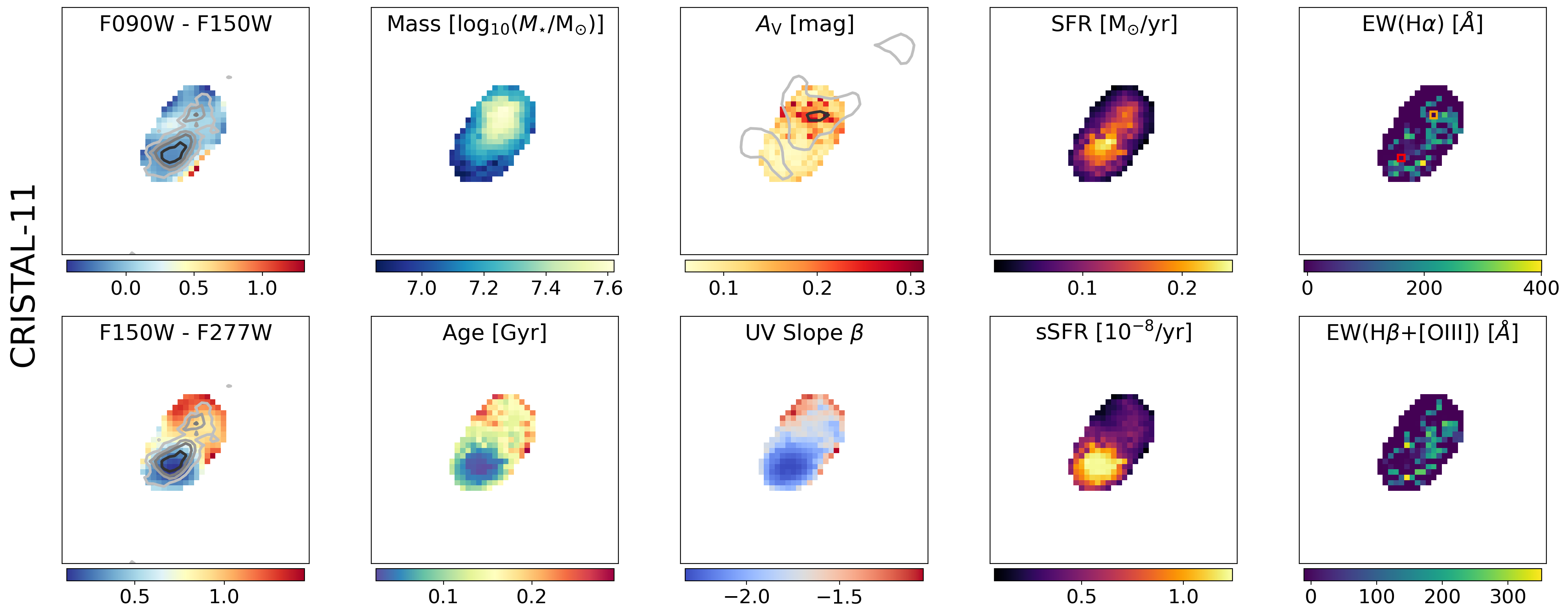}

\vspace{0.5cm}

     \includegraphics[width=0.9\textwidth]{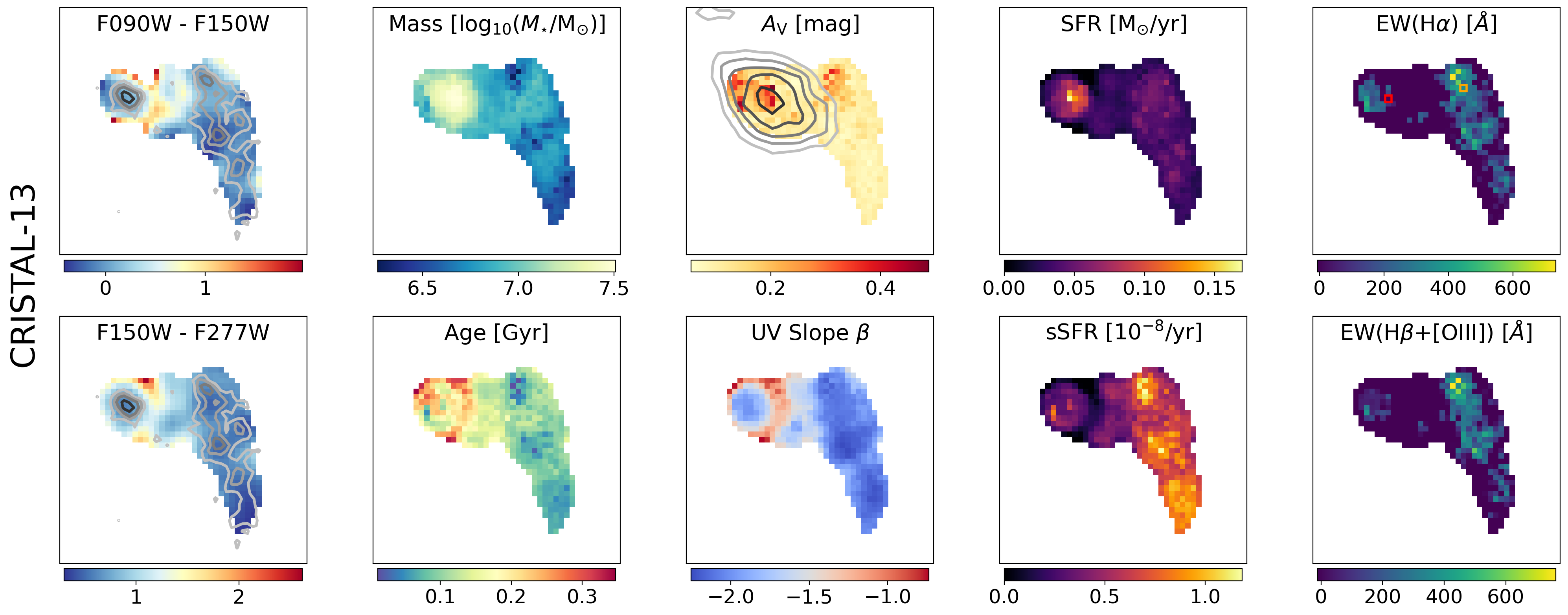}

 \vspace{0.5cm}

    \includegraphics[width=0.9\textwidth]{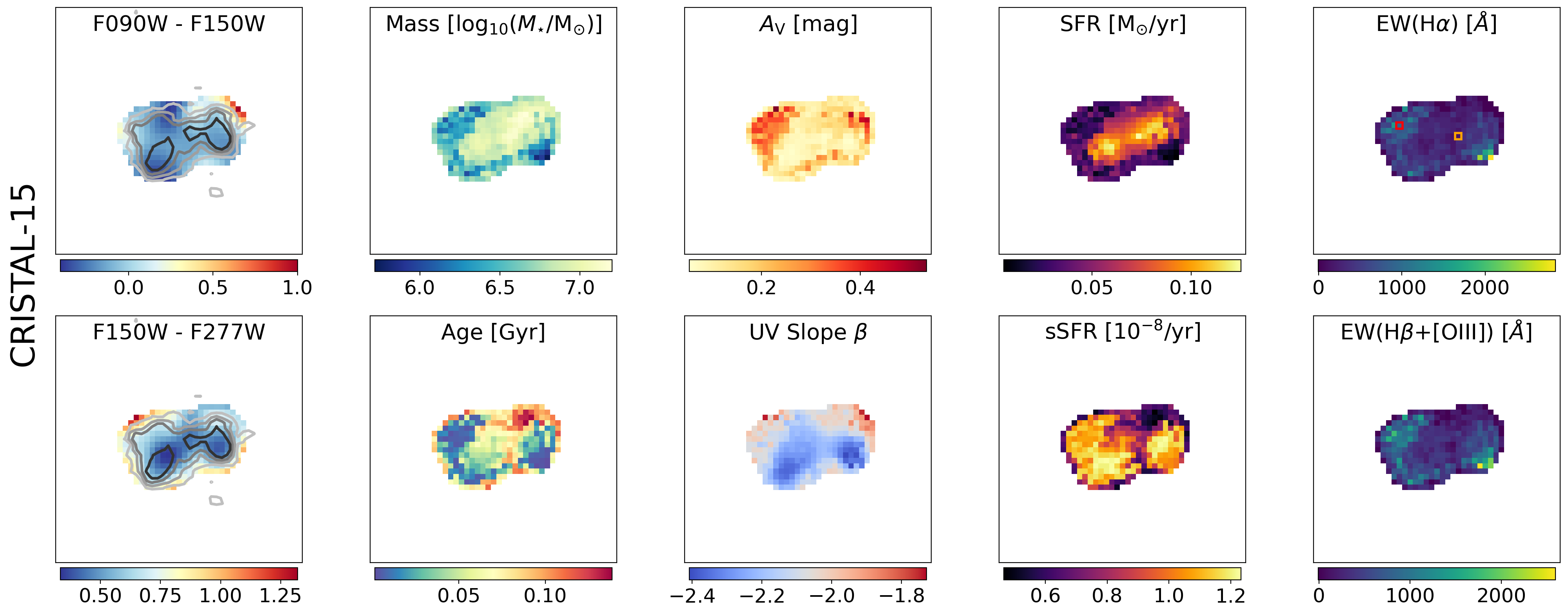}

\caption{The resolved pixel-by-pixel colour maps and SED fitting results of CRISTAL-11, CRISTAL-13 and CRISTAL-15 at $z = 4.45$--$4.58$. 
For each source, the left-most column shows the empirical colour maps; the top image shows the colour across the UV slope given by F090W--F150W, while the bottom shows the colour across that Balmer break given by F150W--F277W. 
The grey/black contours in these colour-colour maps represent the unconvolved F090W filter band at [3, 6, 9, 15]$\sigma$ to guide the eye to the location of the major components.
The other eight images show maps of the parameters, as inferred from the {\tt BAGPIPES} fitting analysis, where units are given per pixel.
On the top row (left to right) we show the logarithm of stellar mass, the dust extinction ($A_{\rm V}$) in magnitudes, the dust corrected SFR and finally the inferred H$\alpha$ rest-frame equivalent width. 
On the bottom row (left to right) we show the age, the rest-frame UV slope $\beta$, the sSFR and the H$\beta +$[OIII] rest-frame equivalent width. 
The contours in the \av~map correspond to the CRISTAL ALMA Band 7 continuum at [3, 4, \dots]$\sigma$ where present. 
Only CRISTAL-11 and CRISTAL-13 have significant detections with ALMA, corresponding to the dust continuum.
Each pixel is 0.03 arcsec, and each image is 1.4 arcsec in width.
We highlight two pixels in the upper-right H$\alpha$ map with red/orange squares.  The corner plots from the SED fitting for these pixels are shown in Appendix~\ref{appendix:pixel}.  Note that the colour bars are different for each source.}
    \label{fig:c11}
\end{figure*}

\begin{figure*}
    \centering
    \vspace{0.5cm}
        \includegraphics[width=0.9\textwidth]{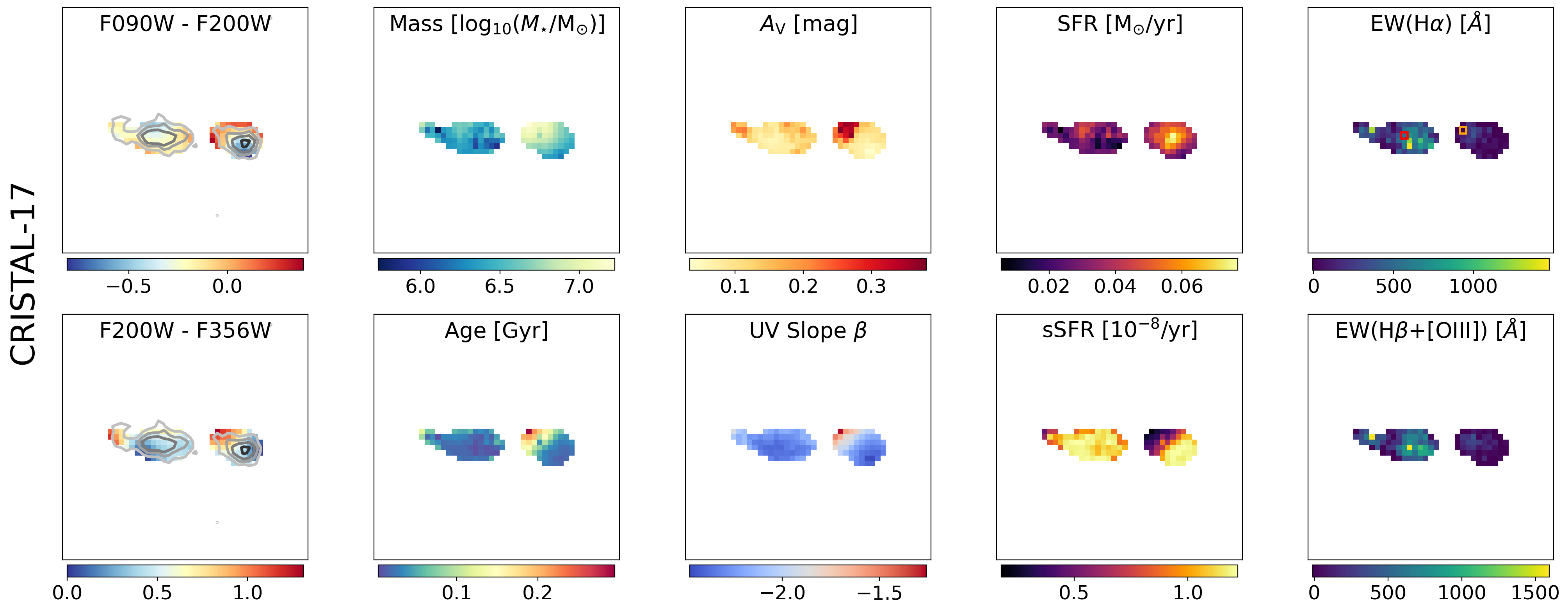}
    \caption{Parameter images and colour maps of CRISTAL-17, the galaxy in our sample that lies at $z = 5.64$. 
    The left-most column of images show the empirical colour-colour maps; the top image shows the colour probing the rest-frame UV slope given by F090W--F200W, while the bottom shows the colour across the Balmer break given by F200W--F356W. 
    The other eight images show maps of the parameters, as described in the caption to Fig.~\ref{fig:c11}, with no contours shown in $A_{\rm V}$ image due to lack of any detection in the ALMA-CRISTAL data above 3$\sigma$ for this galaxy.}
    \label{fig:c17}
\end{figure*}

\section{Results}\label{sect:results}
All four galaxies are clearly spatially resolved in the PRIMER~\emph{JWST} images, extending over several hundred NIRCam pixels at $> 20 \sigma$ significance in the IVW stack, as shown in Fig.~\ref{fig:flux}. 
For all of the galaxies in the sample we are able to identify at least two distinct components at the resolution of~\emph{JWST}.
This is in contrast to the previous~\emph{HST} imaging where the sources typically appears as a single extended object in the rest-frame UV to optical wavelengths.
In this section we first present a detailed exploration of the results of the resolved SED study, followed by a comparison of the derived resolved properties to that from the integrated photometry.

\subsection{Resolved SED fitting} \label{ssec:resolved}
In Fig.~\ref{fig:c11} and Fig.~\ref{fig:c17} we present the pixel-by-pixel fitting results.
These show the {\tt BAGPIPES} derived dust-corrected SFR, stellar mass, specific star-formation rate (sSFR = SFR/{\mstar}), mass-weighted age, and {\av}, in addition to the rest-frame UV slope {$\beta$} and rest-frame equivalent width of the estimated H$\alpha$ and \hboii~emission lines.
We also show empirical colour maps that probe the rest-UV slope and Balmer break to present a model independent view of the galaxy colours in these spectral regions. 
Reassuringly, the process we used to decide whether to include pixels in the fitting analysis provides maps that have a consistent shape to that observed in the original images (Fig.~\ref{fig:flux}).
We include contours from the F090W images prior to PSF homogenisation as an overlay to the empirical colour maps.
We also show dust continuum contours from the ALMA-CRISTAL program in the {\av} image where there was significant (>3{$\sigma$}) signal, in the case of CRISTAL-11 and CRISTAL-13.
Consistent results are also found when assuming a delayed-$\tau$ SFH, as presented in Fig.~\ref{fig:delayed-tau-c11} and Fig.~\ref{fig:delayed-tau-c17}.
In general, for all of the sources we find the empirical colour map that probed the rest-frame UV corresponds well to the SED fitting derived slope {$\beta$}, as expected.
Similarly, the colour probing the Balmer break (shown by the bottom left panels in Fig.~\ref{fig:c11} and Fig.~\ref{fig:c17}) well matches the age and stellar mass maps, giving us further reassurance that the model fitting approach is robust.
Furthermore, younger populations tend to be the bluest in the rest-frame UV slope {$\beta$}, which overlaps with the regions of the highest sSFR and rest-frame optical emission line strength as expected.
We visually inspected the~\emph{HST}/ACS F606W image obtained from the CANDELS survey~\citep{Grogin11,Koekemoer11} for all sources to verify that the different components are drop-out sources, and hence likely at the same redshift.
We now discuss in detail the four sources in turn.

\subsubsection{CRISTAL-11}
The galaxy CRISTAL-11 (DEIMOS\_COSMOS\_630594, at $z=4.44$), appears as a single elongated source in the previous~\emph{HST} optical and NIR data as shown in Fig.~\ref{fig:flux}.
In NIRCam, we find that the source is formed of two distinct components, which we will call North and South in subsequent discussion.
First considering the colour-colour maps for this source, we see that the rest-frame UV colour is fairly uniform and blue (F090W-F150W $\simeq -0.2$) over the two components, while the colour bracketing the Balmer break (F150W$-$F277W) shows a strong gradient of $\Delta {\rm mag} \simeq 1$, with significantly redder colours in the North region.
The derived $\beta$-slope map, in agreement with the empirical colour map of F090W-F150W, shows a relatively uniform blue value over the source, with the North component being slightly redder than the South ($\beta \simeq -1.8$ and $\beta \simeq -2.2$ in the North and South respectively).
These observations suggest that the North region of CRISTAL-11 harbours potentially older stellar populations, or young stars with additional dust reddening.

Using the derived physical properties from the SED fitting analysis we can disentangle these effects and determine the role of age and dust in the observed colours of the source.
The North region, as expected from the redder colours across the Balmer break, contains the peak of the stellar mass, showing a higher dust attenuation (\av $\simeq 0.3\,{\rm mag}$) and an older age ($\simeq 200\,{\rm Myr}$) in comparison to the South region.
Instead, the South component consists of the peak of the star formation in the galaxy, showing evidence for a young ($\simeq 10\,{\rm Myr}$) stellar population that has very little dust attenuation.
This region corresponds to the peak in SFR and sSFR.
If we separate the source by the two regions, we find that the younger South component contains only around a third of the total mass, despite being the brighter component in the~\emph{HST} data. 
We find the inferred \halpha~and \hboii~emission line strengths are relatively weak and the value is affected by the flux errors in the F277W and F356W bands, leading to values that are not strongly correlated with position in the galaxy.
With the new~\emph{JWST} NIRCam data extending to $5\,\mu {\rm m}$ we can extract the centre of mass of the galaxy.
We find that this is offset by approximately 0.5 arcsec (3.5 kpc) from the rest-frame UV peak.
The centre of mass of the system (in the North component) is also the location of the dust continuum detected by ALMA/CRISTAL (discussed further in Section~\ref{sect:disdust}).

\subsubsection{CRISTAL-13}
CRISTAL-13 (vuds\_cosmos\_5100994794, at $z=4.58)$, shown in Fig.~\ref{fig:c11}, has the largest physical extent of the four CRISTAL galaxies studied in this paper.
It is formed of a compact North East component, and a diffuse and clumpy tail to the West, which extends over 1.5 arcsec or 10 kpc.
In the Western tail (WT), we identify at least 5 separate clumps in the higher spatial resolution F090W to F200W bands shown in Fig.~\ref{fig:flux}.
When the data is convolved to the resolution of the F444W band for the SED fitting analysis, these regions are less easily distinguished, however we see structure in the derived parameter maps over these scales (e.g. see the contours shown in the F090W-F150W map in Fig.~\ref{fig:c11}).
The clumps within the Western tail show correlated values of high sSFR, young age, and blue rest-frame UV slope, which are also correlated with the F090W--F150W colour as expected.
We see several peaks in the derived {\sc H}$\alpha$ and \hboii~emission in the Western tail, further corroborating that these are regions of high sSFR with young ($\simeq 50$ Myr) ages.

The NE component and the Western tail are both very blue ($\beta \lesssim -2$) and embedded within regions of redder rest-frame UV colour.
We verified that this halo of redder colours was present in the images directly, and not an artefact of the fitting process.
The NE component is compact and shows a colour gradient such that the galaxy is reddest between the NE component and the blue tail.
This NE regions is characterised by a relatively complex age and \av~structure, which is due to degeneracies between these properties in the fitting.
The redder region of the NE component, which is also picked up in the Balmer (F150W--F277W) colour maps, corresponds to the centre of mass of the system, which is subtly offset from the peak in SFR.
The peak in mass is however co-spatial with the peak in {\av}.
Despite containing a quarter of the pixels, the wider NE component (that includes the SFR and stellar mass peaks) contains half of the total mass. 
CRISTAL-13 contains a significant ALMA detection of the dust continuum (8{$\sigma$}), which spatially coincides with the strongest inferred {\av}. 

\subsubsection{CRISTAL-15}
CRISTAL-15 (vuds\_cosmos\_5101244930, at $z=4.58$), shown in Fig.~\ref{fig:c11}, was initially  observed to have two main components in the~\emph{HST}/ACS imaging.
\emph{JWST}/NIRCam now reveals at least five separate clumps in the short-wavelength bands (e.g. in F200W; Fig~\ref{fig:flux}).
When convolved to the resolution of the F444W filter, several of these clumps are merged leading to a smoother profile, which for the aid of discussion here we split into several regions according to the resulting physical properties.
The two brightest clumps in the short-wavelength images are revealed to be very blue ($\beta \simeq -2.4$) due to their young ages (of a few Myr) and lack of dust attenuation (\av $< 0.1$ mag).
These features have a high sSFR and moderate rest-frame optical emission equivalent widths (\hboii $\simeq 1000$\AA).
The maps of stellar mass and SFR appear to have a subtly different morphology to these two rest-frame UV bright clumps, with the stellar mass and SFR peaking to the North West of the source and not corresponding spatially to either of the rest-UV bright clumps.
We measure a halo of different properties surrounding the central SFR and stellar mass peaks.
The halo shows young ages, high derived \av~and increased rest-frame optical emission line strengths.
Particularly in the region to the NE, which corresponds to two clumps in the F200W imaging, we derive very young ages (of the order of $\sim 10$ Myr) with dust attenuation (\av $\sim 0.4$ mag) and very strong \hboii ($\simeq 1500$\AA).
These regions correspond to high sSFR, moderately dusty star-forming clumps that contribute a negligible fraction of the total SFR of the galaxy.
The radial structure we see could represent an inside-out formation mechanism, however spectroscopic observations will be required to confirm the origin of the observed radial structure.
CRISTAL-15 has no detection of the dust continuum emission from the ALMA-CRISTAL program.

\subsubsection{CRISTAL-17}
CRISTAL-17 (DEIMOS\_COSMOS\_742174, at $z=5.64$), shown in Fig.~\ref{fig:c17} has a higher redshift than the other three galaxies included in this study, and due to having a lower SNR in the PRIMER data it has significantly fewer pixels fitted than the other three galaxies.
In the previous~\emph{HST} imaging it was weakly detected, with only the Eastern component clearly visible.
In NIRCam, we clearly detect and resolve the source into two main clumps, with a weak tail seen to be extending from the Eastern clump.
Both of the main clumps are spatially resolved and appear extended along the East-West direction.
The galaxy is fairly uniform in the majority of the parameter estimates, with only one significantly redder region within the West component.
This region, which we confirm directly in the images does show an increased flux at $\lambda > 4\,\mu {\rm m}$, shows a large dust attenuation (\av $\simeq 0.4$) and corresponds to the high-mass peak in the source. 
As well as hosting the red region and centre of mass of the galaxy, the West clump also contains an extremely blue region ($\beta \simeq -2.5$) with little evidence for significant emission from rest-frame optical lines.
These properties are consistent with a region with a high ionizing photon escape fraction, a possibility that will be further investigated by upcoming~\emph{JWST}/NIRSpec follow-up of the rest-frame optical nebular lines for the CRISTAL sample.
The weak tail to the East of the source shows comparable rest-frame UV colours to the main East component, however shows tentative signs of being redder across the Balmer break, although our analysis is limited by the low SNR of this region.

\begin{table*}
    \centering
    \caption{The physical properties of the four galaxies studied in this work.  We show estimates of the SFR, stellar mass, age, \av, rest-frame UV slope $\beta$, and the \halpha~equivalent width as inferred from {\tt BAGPIPES} assuming a CSFH.
    Each galaxy is shown as a separate column.
    For each parameter, the three rows correspond to the resolved (pixel-by-pixel) analysis, followed by the integrated analysis. 
    Errors shown correspond to the 16th and 84th percentile estimations of the best-fit parameters, and the resolved results correspond to either a sum (for the SFR, stellar mass) or an average (age, \av, $\beta$, ${\rm EW}_{\rm 0}$).
    The final row for each parameter shows the previous values determined as part of the ALPINE program, as described in~\citet{Bethermin20} and~\citet{Faisst20}.
    The \halpha~rest-frame equivalent width is the measured value as derived from the {\tt BAGPIPES} best-fitting SED model.
    For the ALPINE results we have reversed the dust correction applied in~\citet{Faisst20} so these values can be compared directly.
    }
    \label{tab:params}
\begin{tabular}{ll|cccc}
\hline
                             &            & CRISTAL-11                                            & CRISTAL-13                                            & CRISTAL-15                                            & CRISTAL-17                                            \\ \hline
                             & Resolved   & $26.8\substack{+12.3\\  -6.7}$& $24.7\substack{+13.1\\ -10.1}$& $20.1\substack{+6.7\\  -4.8}$& $8.4\substack{+4.0\\  -2.7}$\\[1ex]
SFR {[\sfrunit]}             & Integrated & $19.0\substack{+9.4\\  -4.1}$& $32.5\substack{+4.2\\  -5.4}$& $15.3\substack{+6.2\\  -3.3}$& $2.7\substack{+1.1\\  -0.5}$\\[1ex]
                             & ALPINE     & $31.5\substack{+23.9 \\ -15.3}$                       & $28.4\substack{+22.8 \\ -14.0}$                       & $26.1\substack{+20.0 \\ -10.6}$                       & $12.9\substack{+8.9 \\ -3.3}$                         \\ \hline
                             & Resolved   & $9.71\substack{+0.08\\  -0.08}$& $9.77\substack{+0.11\\  -0.10}$& $9.36\substack{+0.10\\  -0.12}$& $9.02\substack{+0.14\\  -0.16}$\\[1ex]
Mass {[\lmstar]}             & Integrated & $9.69\substack{+0.07\\  -0.08}$& $9.49\substack{+0.08\\  -0.16}$& $9.12\substack{+0.14\\  -0.11}$& $8.42\substack{+0.12\\  -0.11}$\\[1ex]
                             & ALPINE     & $9.77\substack{+0.14 \\ -0.15}$                       & $9.73\substack{+0.15 \\ -0.13}$                       & $9.67\substack{+0.16 \\ -0.13}$                       & $9.56\substack{+0.13 \\ -0.15}$                       \\ \hline
                             & Resolved   & $128\substack{+118\\  -54}$& $128\substack{+107\\  -54}$& $53\substack{+45\\  -24}$& $50\substack{+53\\  -25}$\\[1ex]
Mass-weighted age {[Myr]}                  & Integrated & $183\substack{+88\\  -96}$& $50\substack{+40\\  -17}$& $17\substack{+16\\  -6}$& $7\substack{+4\\  -2}$\\[1ex]
                             & ALPINE     & $186\substack{+227 \\ -95}$                           & $193\substack{+252 \\ -95}$                           & $183\substack{+364 \\ -88}$                           & $354\substack{+338 \\ -215}$                          \\ \hline
        & Resolved   & $0.123\substack{+0.148\\  -0.088}$& $0.129\substack{+0.147\\  -0.089}$& $0.164\substack{+0.111\\  -0.086}$& $0.127\substack{+0.117\\  -0.084}$\\[1ex]
$A_{\rm V} [\rm mag]$                   & Integrated & $0.181\substack{+0.143 \\ -0.121}$                    & $0.362\substack{+0.027 \\ -0.057}$                    & $0.173\substack{+0.083 \\ -0.078}$                    & $0.123\substack{+0.088 \\ -0.077}$                    \\[1ex]
                             & ALPINE     & $0.8$                                                 & $0.6$                                                 & $0.4$                                                 & $0.0$                                                 \\ \hline
                             & Resolved   & $-1.84\substack{+0.04\\  -0.04}$& $-1.82\substack{+0.04\\  -0.04}$& $-2.15\substack{+0.03\\  -0.03}$& $-2.20\substack{+0.03\\  -0.03}$\\[1ex]
Rest-frame UV slope, $\beta$ & Integrated & $-2.10\substack{+0.02 \\ -0.02}$                      & $-1.88\substack{+0.02 \\ -0.02}$                      & $-2.21\substack{+0.02 \\ -0.02}$                      & $-2.39\substack{+0.02 \\ -0.02}$                      \\[1ex]
                             & ALPINE     & $-1.44\substack{+0.30 \\ -0.24}$                      & $-1.63\substack{+0.20 \\ -0.17}$                      & $-1.87\substack{+0.14 \\ -0.17}$                      & $-2.13\substack{+0.34 \\ -0.23}$                      \\ \hline
                             & Resolved   & $53\substack{+42\\  -53}$& $89\substack{+72 \\ -72}$& $423\substack{+101 \\ -101}$& $266\substack{+191\\  -191}$\\[1ex]
{\sc H}$\alpha\, {\rm EW}_{0}$[\AA] & Integrated & $247 \pm 98$& $396 \pm 67$& $572\pm 107$& $820\pm175$\\[1ex]
                            & ALPINE     & $670\substack{+350 \\ -220}$                         & $580\substack{+220 \\ -160}$                         & $420\substack{+160 \\ -120}$                          & --   \\ \hline   

\end{tabular}

\end{table*}

\subsection{Resolved and integrated physical properties} \label{ssec:integrated}

\begin{figure*}
    \centering
    \includegraphics[width = 0.49\linewidth]{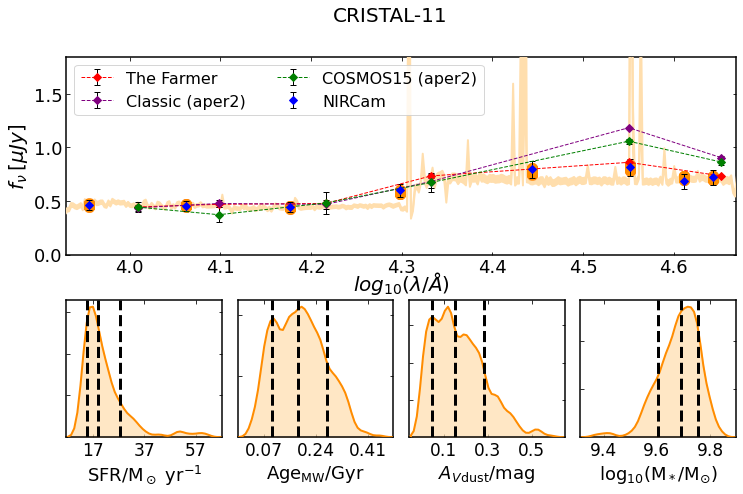}
       \includegraphics[width = 0.49\linewidth]{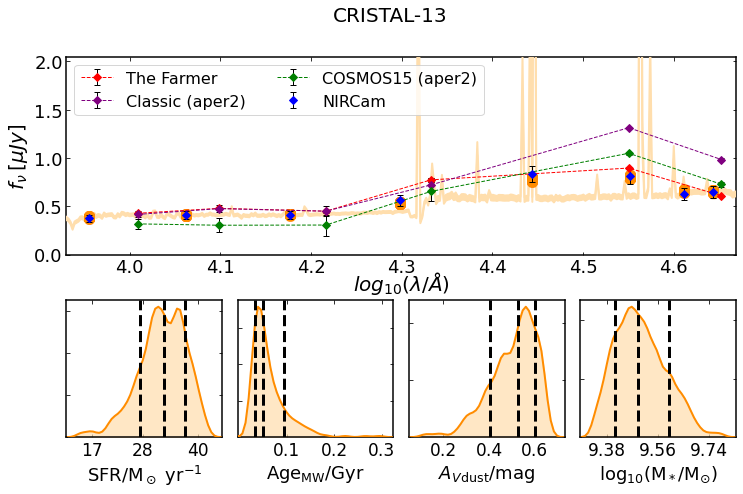}\\
        \includegraphics[width = 0.49\linewidth]{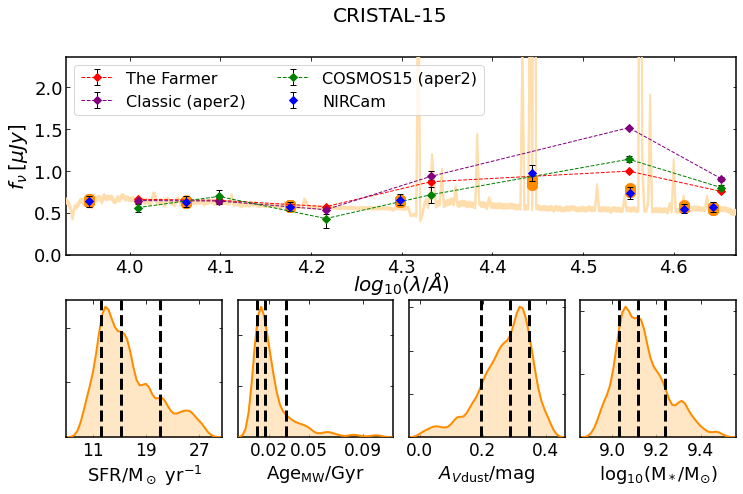}
       \includegraphics[width = 0.49\linewidth]{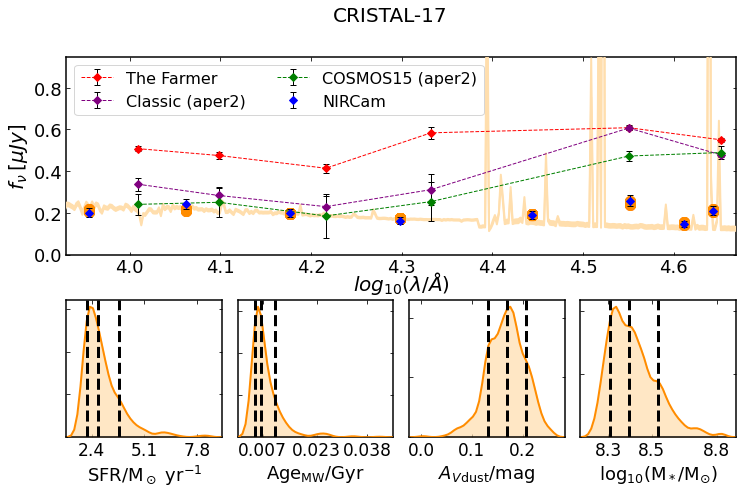}\\
    \caption{The integrated photometry, best-fitting SED models and fitted parameter estimates for the four CRISTAL galaxies studied in this work. 
    Points shown in blue are the extracted fluxes from NIRCam summed over the pixels that satisfy the threshold as specified in section \ref{ssec:setup}. 
    The posterior SED models fitted to this data by {\tt BAGPIPES} are shown in orange, shaded from the 16th to the 84th percentile. 
    Flux values from the COSMOS2020 catalogue for the UltraVISTA ($YJHK_{\rm s}$) and~\emph{Spitzer}/IRAC \chone~and \chtwo~bands, extracted using {\tt SExtractor} ({\tt THE FARMER}), are shown in purple (red). 
    {\tt SExtractor} derived aperture flux values from the COSMOS2015 catalogue are shown in green.
    The four subplots for each galaxy show the best-fit estimation of SFR, mass-weighted age, $A_{\rm V}$, and the logarithm of the stellar mass, with the 16th, 50th and 84th percentile estimates indicated as vertical dashed lines.}
    \label{fig:integrated-sed}
\end{figure*}
   
The resulting physical parameters from the resolved and integrated fitting methods are presented in Table \ref{tab:params}. 
We also present the consistent results from Li et al. (2024) in Appendix~\ref{appendix:li} for comparison.
The integrated fits, which are shown in Fig.~\ref{fig:integrated-sed}, had reduced {$\chi^2$} values of 0.6, 0.8, 1.3 and 1.4 for CRISTAL-11, -13, -15 and -17 respectively, and the posterior distribution seems to indicate a good fit. 
The integrated fits assuming a delayed-$\tau$ model are shown in Appendix~\ref{appendix:sfh} in Fig.~\ref{fig:delayed_integrated SED}.
The reduced {$\chi^2$} values for the delayed-$\tau$ model are comparable with 0.5, 0.6, 1.1 and 1.4.
The integrated and resolved estimates of the total SFR as derived from SED fitting with a CSFH are consistent within the error range for all four galaxies. 
We also find good consistency between the stellar masses, which are within 0.1--0.3 dex at $z \simeq 4.5$ without any evidence for the resolved values being increased by $> 0.5{\rm ~dex}$ as found by~\citet{GimenezArteaga23}.
In CRISTAL-17 we find a larger offset, such that the resolved mass is 0.6 dex lower than the integrated value. 
The comparison between our integrated and resolved values follow the mass trends found in previous studies (e.g. \cite{GimenezArteaga23, PerezGonzalez23, Shen24}; see Figure \ref{fig:intmass}, and discussion in Section~\ref{sect:discussion}).
Comparing the mass-weighted ages, we find that these tend to be higher in the resolved case than from the integrated single fit, however here we are comparing the best-fit age to the average age and hence the comparison is not straightfoward.
We also find that the lower mass galaxies (CRISTAL-15, CRISTAL-17) show lower ages of around $50\,{\rm Myr}$, while the CRISTAL-11 and CRISTAL-13 values are more than twice this ($\simeq 130\,{\rm Myr}$) as derived as the mean of the pixel-by-pixel fitting.
This difference across the sample is also reflected in the resolved $\beta$-slopes and \av, which show that CRISTAL-11 and CRISTAL-13 (the older galaxies) are slightly redder.
In comparison to the main-sequence (MS) of star formation, the four sources with the fiducial SED fitting properties derived for the sample in Herrera-Camus et al. 2025 lay within $\sim 0.3{\rm dex}$ of the relation at $z = 5$ from~\citet{Speagle14}.
With the resolved fitting results presented here we find no significant differences for CRISTAL-11 and CRISTAL-13, such that their position on the MS is effectively unchanged.
For CRISTAL-15 and CRISTAL-17 we recover a lower \mstar~in this analysis, however as the recovered SFRs are also lower the sources are still consistent with the $z = 5$ MS.

\subsubsection{Comparison to ALPINE}
We compare our new~\emph{JWST}/NIRCam based SED fitting results to the original ALPINE values presented in~\citet{Faisst20} and~\citet{Bethermin20}.
The SED fitting analysis in this case was performed using {\sc Le Phare} assuming a range of SFHs including exponentially declining, and the delayed exponential ($\tau$) models and CSFH that we use in this work.
\citet{Faisst20} assume the same initial mass function (IMF,~\citet{Chabrier03}) and dust law~\citep{Calzetti00} as we do, although they fix the metallicity to be $[0.2, 1.0]\,{\rm Z}_{\odot}$.
The minimum age allowed in their model is $50\,{\rm Myr}$, and they include emission lines according to the default {\sc Le Phare} prescription.
As shown in Table~\ref{tab:params}, we find several differences between the fitting results from ALPINE (based on the COSMOS2015 catalogue;~\citealp{Laigle16}) and our analysis.
Comparing the integrated fit results from our analysis, the SFRs are lower than the previous ALPINE analysis by $3$--$12~$\Msun/yr.
Similarly, our stellar masses are lower than the ALPINE values by $\simeq 0.1 (0.2)\,{\rm dex}$ for the older/more massive galaxies CRISTAL-11 (CRISTAL-13), and $0.55\,{\rm dex}$ and $1.1\,{\rm dex}$ for CRISTAL-15 and CRISTAL-17 respectively.
The significant offset between the values for CRISTAL-17 is likely due to blending with a nearby source, as we discuss in Section~\ref{sect:phot}.
The potential origin of the other differences can be either due to photometric differences, which we investigate in Fig.~\ref{fig:integrated-sed}, and/or due to the different SED fitting parametrisations.
The ALPINE fits tend to show older ages than both our resolved and integrated NIRCam based fits.
Noticeably, we conclude that CRISTAL-17 is the youngest of all four galaxies with the resolved and integrated predictions being $50\substack{+53 \\ -25}$ and $7\substack{+4 \\ -2}$ Myr respectively, while the original catalogue taken from the ALPINE survey has this source as the oldest with an age of $354\substack{+338 \\ -215}$ Myr amongst the CRISTAL galaxies. 
With the exception of CRISTAL-17, the ALPINE fits all prefer a larger dust attenuation.
Hence, assuming that the photometric measurements were comparable, the larger dust attenuation and older ages that are best-fit in the ALPINE analysis can explain the observed higher stellar masses and SFRs derived in this previous analysis.
Turning now to the rest-frame UV slope, which was measured via a power law fit to the SED model in~\citet{Faisst20}, we find bluer slopes in our NIRCam analysis for all of the sources.
We recover slightly lower observed \halpha~${\rm EW}_{0} \simeq 300$--$400$\AA~for CRISTAL-11 and CRISTAL-13, and estimate a higher value for CRISTAL-15 of ${\rm EW}_{0} \simeq 600$\AA~(although we find agreement within the large errors for C13 and C15 to the previous results).
No $EW_{\rm 0}$ was reported for CRISTAL-17 in the ALPINE analysis.
The \halpha~values in~\citet{Faisst20} were derived using the method in~\citet{Faisst19} applied to the 3 arcsec aperture photometry from the COSMOS2015 catalogue.
They were corrected for dust attenuation assuming the best-fit from {\sc Le Phare} and a differential stellar to nebular attenuation of $E_{\rm s}(B-V)/E_{\rm n}(B-V) = 0.44$, however we reversed this correction for a valid comparison here.

\subsubsection{Comparison to COSMOS2015 and COSMOS2020 }\label{sect:phot}
We investigated the differences in the SED fitting results from previous analyses further by directly comparing the photometry from the COSMOS2015 and COSMOS2020 catalogues to the fluxes derived from~\emph{JWST}/NIRCam in Fig.~\ref{fig:integrated-sed}.
We show the flux values from the four UltraVISTA NIR bands ($YJHK_s$) and the two~\emph{Spitzer}/IRAC warm channels, \chone~and \chtwo.
This data was obtained from the catalogues presented in ~\citet{Laigle16} and~\citet{Weaver22} and we show photometry measured using two different pieces of software. 
The {\tt Source Extractor} \citep{Bertin96} values were obtained using fixed 2 arcsec diameter apertures (corrected to total assuming a point source correction), while {\tt THE FARMER} (only for COSMOS2020;~\citealp{Weaver23}) results were obtained using model fitting to the images (and thus represent a total flux measure).
The COSMOS2015 and COSMOS2020 aperture values are roughly consistent, with some small offsets (e.g. in CRISTAL-13) but the same trends with wavelength.  In the following discussion we therefore refer only to the COSMOS2020 values for conciseness.
We find that our measured~\emph{JWST} fluxes are consistent with the UltraVISTA $YJH$ photometry probing the rest-frame UV for the sources, but are noticeably lower than the COSMOS2020 values in the $K_{\rm s}$, \chone~and/or~\chtwo~bands in all sources.
In general, the model-based {\tt THE FARMER} and aperture-based {\tt Source Extractor} photometry match in the rest-frame UV region, but again deviate from each other at longer wavelengths, particularly in \chone.
CRISTAL-17 has the most discrepant photometry between the two COSMOS2020 results, with NIRCam fluxes upwards of 1 magnitude fainter than the COSMOS2020 catalogues as derived from {\tt THE FARMER}.
We identified a lower-redshift companion (detected in the optical) at 1.1 arcsec to the North of CRISTAL-17, which is contaminating the flux for CRISTAL-17 and leading to an overestimate of the properties.
The observed offsets between the previous best and new NIRCam photometry explains in particular the slightly reduced \halpha~strengths that we derived from the multi-band~\emph{JWST} imaging for CRISTAL-11 and CRISTAL-13.
As shown in Fig.~\ref{fig:integrated-sed}, the \chone~results are typically higher in the aperture photometry case, suggesting that confusion in the IRAC bands is the cause of the discrepancy.
Interestingly,~\citet{Faisst20} find that the dust-corrected \halpha~emission line strength of the full ALPINE sample (of which CRISTAL represents the higher mass end) was in excess of that expected given the stellar mass from lower redshift relations.
While this is expected for higher-redshift galaxies due to the higher sSFR (and normalisation of the star-forming main sequence), the reduction in the flux around \halpha~in the new~\emph{JWST} data at $\lambda = 3$--$5\mu{\rm m}$ hints that the \halpha~emission line strengths may be weaker than previously thought.

\section{Discussion}\label{sect:discussion}
In this work we have compared data from the ALMA-CRISTAL program, which provides high-resolution observations of the dust continuum, to deep~\emph{JWST}/NIRCam imaging from PRIMER to investigate in detail the properties of four $z \simeq 5$ galaxies.

\subsection{The rest-UV and optical morphology}
As shown in Fig.~\ref{fig:flux}, the improvement in sensitivity and spatial resolution provided by the~\emph{JWST} PRIMER imaging compared to the previous~\emph{HST} data is dramatic.
In common with many other studies of high-redshift galaxies (e.g.~\citealp{Barisic17, Bowler17, Boyett24}), we find that all four galaxies show multiple components in the imaging probing the rest-frame UV.
This observation is in agreement with the observed rise with redshift in the proportion of galaxies that show a clumpy or irregular morphology (e.g.~\citealp{Guo18, Sok22}).
We find between $2$--$8$ separate clumps at the resolution of our data (with a full-width-half-maximum (FWHM) of the F200W band, $0.06\,$arcsec, corresponding to $\sim 400$ pc).
Crucially however, with~\emph{JWST} we can probe the observed wavelengths at $\lambda \gtrsim 2\,\mu{\rm m}$ providing the first measurement of the rest-frame optical morphology of these galaxies, a regime that was previously only accessible with unresolved~\emph{Spitzer}/IRAC imaging (e.g. Fig.~\ref{fig:integrated-sed}).
It has been argued that high-redshift galaxies only appear clumpy due to the majority being selected based on the rest-frame UV light, a regime which is in general observed to be fractured even in lower-redshift galaxies (e.g.~\citealp{Elmegreen09, Wuyts12}).
Recent studies of individual high-redshift sources with~\emph{JWST} in the rest-frame optical have seen evidence for evolved disks, which have a smoother and/or more centrally concentrated morphology in comparison to the rest-frame UV.
For example, \citet{Nelson23} identified a mature rotating disk in a $z = 5.3$ galaxy within the spectroscopic FRESCO survey.
`Twister-z5' is a massive (\lmstar $\simeq 10.4 \pm 0.4$) galaxy that has centrally depressed \halpha~emission and sSFR, showing tentative evidence for inside-out growth and early bulge formation.
Similarly,~\citet{Setton24} found a compact red core in the comparably massive quiescent galaxy at $z = 3.97$, with the core appearing to be older and dustier than the outskirts.

While these examples are particularly massive, the recent detection of the `Cosmic Grapes' source at $z = 6.072$ provides evidence that disk-like structures may exist at lower masses.
\citet{Fujimoto24} identified 15 individual clumps in the \lmstar $\simeq 8.7$ source that is multiply imaged by the foreground cluster RXCJ0600-2007.
They suggest that the clumps are embedded within a rotating disk structure (albeit a dynamically unstable one), with the stars between the clumps showing older ages~\citep{GimenezArteaga24}.
\citet{GimenezArteaga24} further demonstrate that outshining is a significant problem for this source, with the stellar mass underestimated by up to 0.5 dex using the integrated photometry and assuming a CSFH.

In contrast to these studies, we find a consistent rest-frame UV and optical morphology in the CRISTAL galaxies we study.
This can be seen in the raw data (Fig.~\ref{fig:flux}) and also in the comparison between the \mstar~and SFR maps (Fig.~\ref{fig:c11} and Fig.~\ref{fig:c17}).
The exception is CRISTAL-11, which does show an offset between the peak of \mstar~and SFR, however the overall morphology across the different wavelengths is similar in this case.
We see gradients in the colour in our pixel-by-pixel fits, which are also present in the derived physical properties, however the morphology of the galaxies with wavelength is very similar and we see no strong evidence for older underlying disk morphologies.
As we discuss further in the next sections, this observation, in comparison to simulations, supports a merger origin for the CRISTAL galaxies we study in this work.

\subsection{The rest-UV and optical sizes}
Another way to probe the morphology of old and young stellar light is to compare the observed sizes of galaxies above and below the Balmer break.
\citet{Ono24} and~\citet{Morishita24} showed that the ratio of the rest-optical to UV size was approximately unity in a sample of galaxies from $z = 4$--$10$ and $z = 5$--$14$ respectively.
\citet{Ono24} take this observation as evidence that these galaxies are being caught at the start of inside-out growth, where the young and old stars have yet to be spatially differentiated.
A key complication in this comparison comes from the complex and diverse morphologies of the CRISTAL galaxies (and high-redshift galaxies in general).
Indeed, in the analysis of~\citet{Varadaraj24}, all four of the galaxies we study here were present in the parent sample but failed in the S\'{e}rsic fitting step due to poor residuals.
It is apparent in our analysis that the CRISTAL galaxies cannot be adequately described with a single S\'{e}rsic model, and hence we quote the non-parametric sizes from~\citet{Varadaraj24} instead.
As shown in Table~\ref{tab:fir}, we find that the rest-frame UV and optical sizes are consistent within the errors, with the exception of CRISTAL-13.
CRISTAL-13 has the most extended and complex morphology of our sample, and also potentially harbours a disk like feature in the WT (although this is blue/young and does not represent the centre of mass of the system).
The sizes of the CRISTAL sources are typical for their mass, with half-light radii of around $1\,{\rm kpc}$.
Hence we see no evidence for different rest-frame UV and optical sizes, that might indicate inside-out growth being established or multiple stellar population ages that are significantly spatially decoupled.
As we have previously discussed, the half-light radius does not fully describe the extent of the galaxy, which is of the order of $5$--$10\,{\rm kpc}$ at \lmstar$\simeq 9.5$ as we have shown in Fig.~\ref{fig:flux}.
The size comparison here further supports the visually identified similar morphology across the broad observed wavelength range probed by~\emph{JWST} NIRCam.
We discuss further in Section~\ref{sect:disdust} the differences in size and morphology between the rest-frame UV/optical light and the observed FIR dust continuum emission.

\subsection{A lack of outshining in the CRISTAL galaxies}

\begin{figure}
	\includegraphics[width=\columnwidth, trim = 0.0cm 0.0cm 0.5cm 2cm]{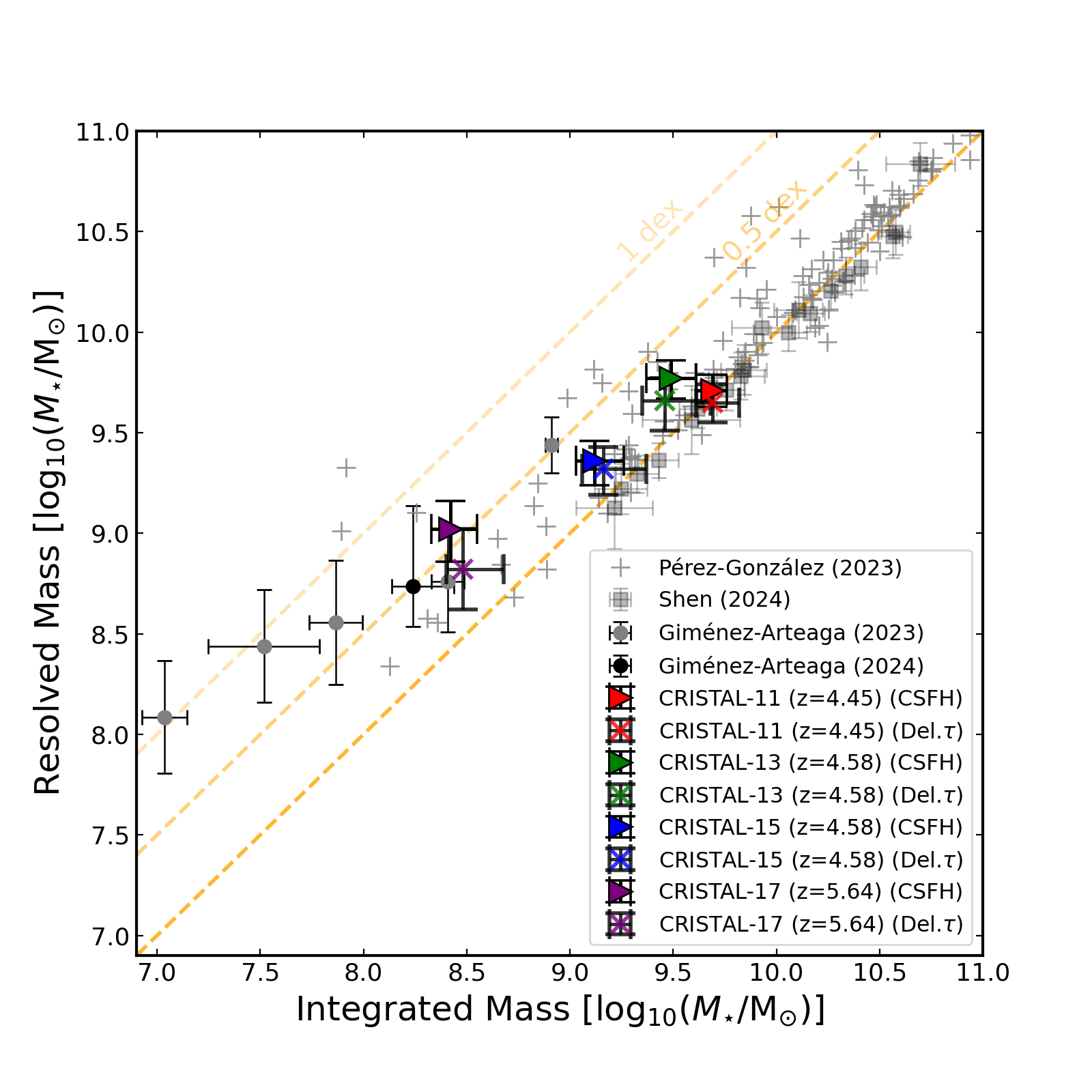}
    \caption{ The integrated \mstar~shown in comparison to the resolved value as derived from our SED fitting analysis.
    The integrated mass is computed from the integrated photometry, while the resolved is determined by summing the stellar mass from each individually fit pixel.  
    The fiducial results assuming a CSFH are shown as the triangles, with the results assuming a delayed-$\tau$ SFH shown for comparison as stars.
    We compare to the study of~\citet{GimenezArteaga23} at $z = 5$--$9$, \citet{Shen24} at $z = 0.6$--$2.2$~\citet{PerezGonzalez23} at $z = 2$--$6$, and show the $z = 6.0$ galaxy `Cosmic Grapes' from~\citet{GimenezArteaga24}.
    Lines of equal integrated and resolved stellar mass, 0.5 dex offset and 1 dex offset are shown.}
    \label{fig:intmass}
\end{figure}

With our resolved SED fitting analysis using the PRIMER data, we can test the impact of outshining in galaxies of higher stellar mass, which arguably could be expected to show a more established stellar population and hence a smaller impact of extremely young bursts that leads to outshining.
We compare the stellar masses from the resolved and integrated fits in Fig.~\ref{fig:intmass}.
Here we show the results from our fiducial analysis, which assumed a CSFH, and the results from assuming a delayed-$\tau$ SFH for comparison.
The results are consistent between these two different SFHs, providing reassurance in our results and allowing a more direct comparison to previous studies (who typically used one of these parameterizations).
We compare to the five galaxies studied by~\citet{GimenezArteaga23}, which were gravitationally lensed (we show the delensed masses), and to the sample of star-forming galaxies at $z = 0.2$--$2.5$ by~\citet{Shen24}.
\citet{Shen24} analyse galaxies with \lmstar$> 9$ using a resolved and integrated approach, assuming a delayed exponential SFH.
We also compare to the red galaxies studied by~\citet{PerezGonzalez23} at $z =2$--$6$ who assumed a delayed-$\tau$ SFH.
We find that in the three CRISTAL galaxies we study with \lmstar$> 9$ we recover a higher mass in the resolved SED fitting analysis, but only by $\simeq 0.1$--$0.3\,{\rm dex}$.
For CRISTAL-17, the lowest mass (and highest redshift) source in our sample, we see a larger deviation.
Hence we are able to bridge between the studies of~\citet{GimenezArteaga23} at \lmstar $< 9$ and the higher mass studies of~\citet{Shen24} and~\citet{PerezGonzalez23}, showing the onset of outshining at the lower mass end.
Our conclusions are unchanged if we instead assume a delayed-$\tau$ SFH, with the recovered stellar masses being within the errors of the CSFH fiducial case.
Although in general the delayed-$\tau$ models produce closer resolved and integrated \mstar~ estimates.

In comparison to~\citet{GimenezArteaga23}, we find that the mass-weighted stellar ages of our galaxies in the integrated analysis are significantly older (50-130 Myrs vs. the $\simeq 2 \,{\rm Myr}$ in that study) and crucially the derived integrated ages are not dramatically different from the average resolved mass-weighted age.
This results in comparable SFRs and \mstar~values between the resolved and integrated fits for the CRISTAL galaxies.
As shown in Fig.~\ref{fig:intmass} our results, coupled with previous studies, show that at \lmstar $\gtrsim 9$ the resolved and integrated stellar masses are in good agreement, whereas at lower masses the effect of outshining can be seen and the two values diverge (e.g. for CRISTAL-17).
We find lower rest-frame optical emission line equivalent-widths than the~\citet{GimenezArteaga23} galaxies, which is directly related to the older ages (than the lower-mass lensed sources studied in~\citealp{GimenezArteaga23, GimenezArteaga24, Fujimoto24}) we recover in general for the CRISTAL sources.
Even for individual clumps, we do not find any regions that exceed \ew (\hboii) $\simeq 1000$\AA~(with the possible exception of CRISTAL-15), however~\citet{GimenezArteaga23} found regions with \ew (\hboii)$> 3000$\AA.
Future spectroscopic observations of the CRISTAL sample with the~\emph{JWST} NIRSpec IFU will provide resolved emission line strengths and thus a more detailed analysis of the impact of these lines on the photometry and refined estimates of the stellar ages.

\subsection{The resolved IRX-$\boldsymbol{\beta}$ relation}\label{sect:disdust}

The presence of dust in the CRISTAL galaxies can be ascertained in two ways, first through the observation of reddening of the rest-frame UV light and second with the detection of emission directly from the dust in the FIR.
At low redshift the energy balance between the \Lir~emanating from the dust with the absorbed \Luv~from young stars has been shown to lead to a correlation between the ratio of the FIR luminosity to the UV luminosity (infrared-excess $=$ {\irx} $= \log_{10}[$\Lir/\Luv]) and its rest-frame UV slope $\beta$ (e.g. the seminal work by~\citealp{Meurer99, Calzetti00}).  
Targeted observations such as the ALMA-ALPINE and ALMA-REBELS large programs have obtained measurements of the infrared luminosity for samples of high-redshift galaxies, enabling the investigation of the evolution of the {\irxb} relation with properties such as redshift and stellar mass \citep[e.g.][]{Fudamoto20a, Fudamoto20, Bowler18, Bowler24}.  
With the improved spatial resolution now available in both the rest-frame UV and FIR with~\emph{JWST} and ALMA, it is now possible to investigate if this relation holds on spatially resolved scales and understand how the distribution of dust within galaxies affects global measurements.  
This may help explain the low {\irx} values of galaxies such as those at $5.1<z<5.7$ studied by~\cite{Capak15}, who concluded that the dust properties were evolving with redshift (e.g. to a lower-metallicity, Small Magellanic Cloud - SMC-like dust).  
Spatially resolved analysis can potentially reveal whether the position of these galaxies on the {\irxb} diagram is due to dust temperature increasing with redshift or geometric offsets between the stars and dust within the galaxy~\citep{Barisic17, Faisst17, Bowler22}. 

\begin{figure}
	\includegraphics[width=0.5\columnwidth]{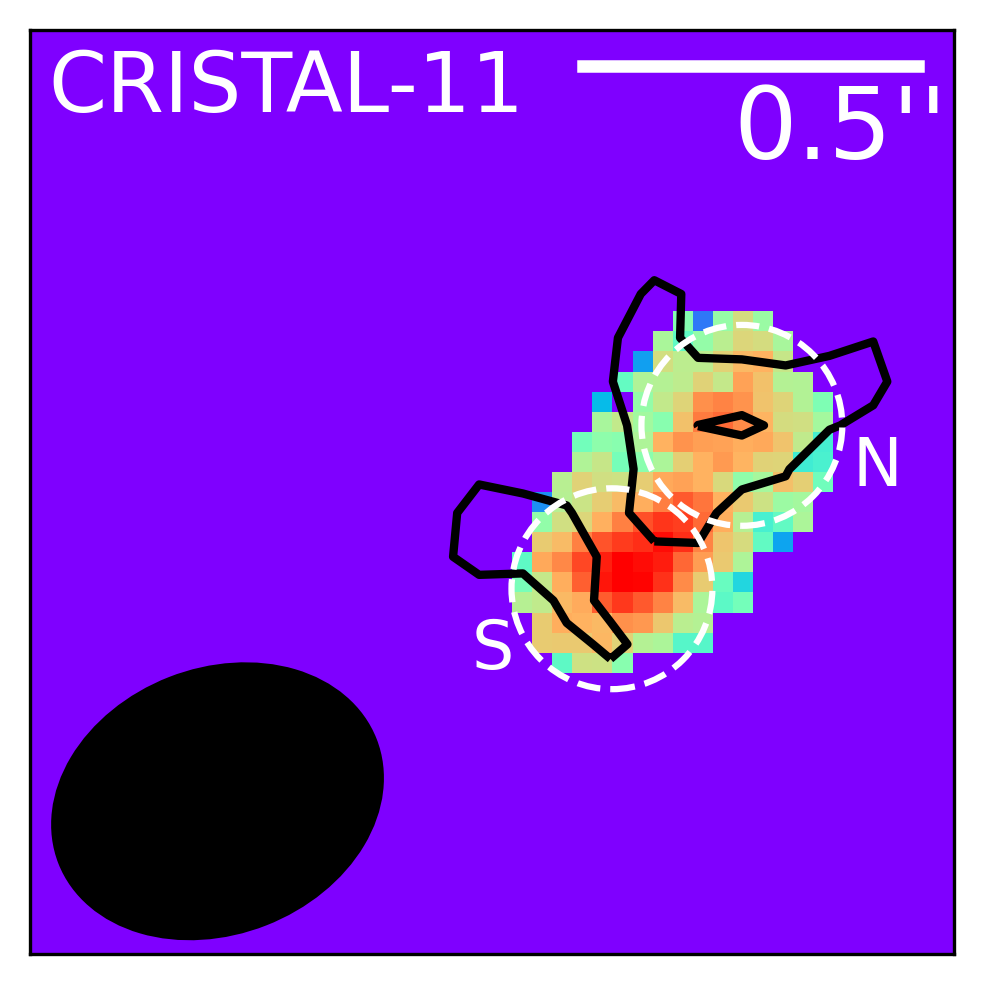}
    \includegraphics[width=0.5\columnwidth]{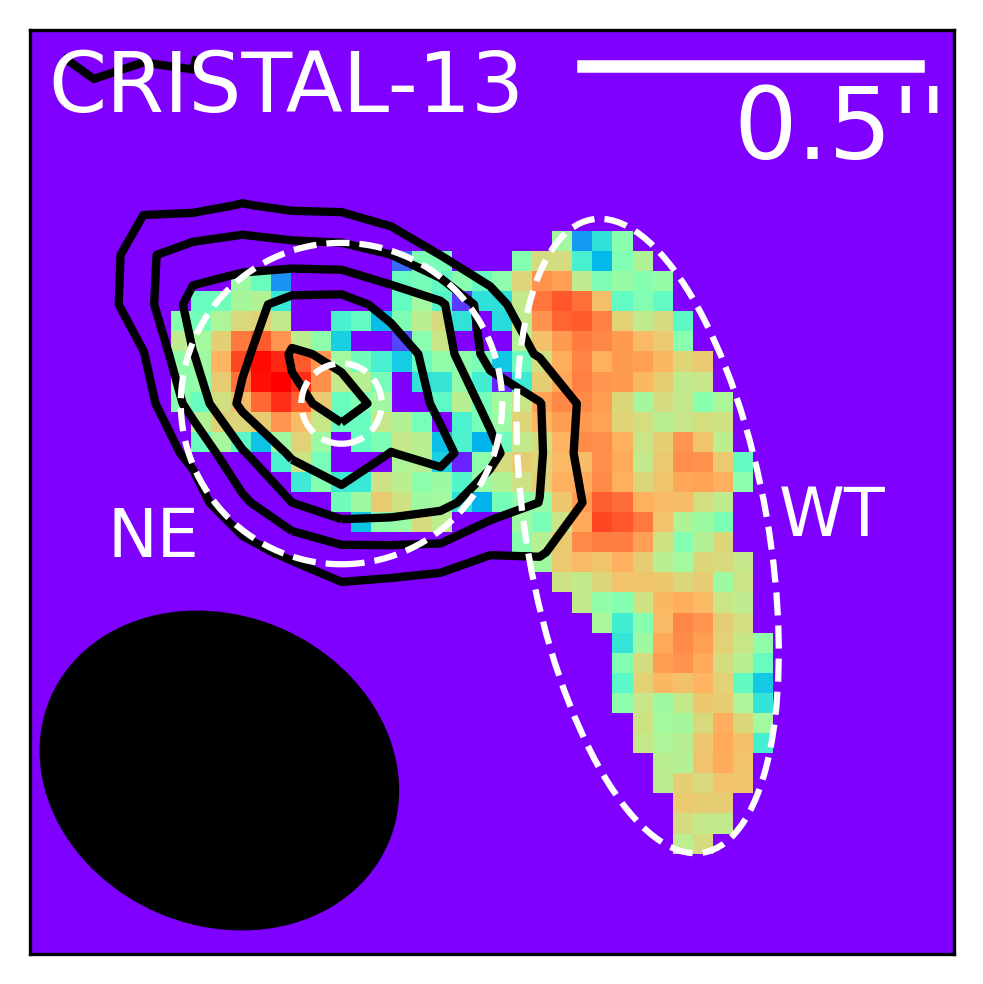}
    \caption{\emph{JWST} F090W filter band image stamps scaled in magnitude space between the peak surface brightness and 28 mag/arcsec$^2$. The stamps are 1.4 arcsec on a side, with North to the top and East to the left. 
    The dust continuum CRISTAL/ALMA Band 7 contours in $1\sigma$ intervals starting at $3\sigma$ are shown by the solid black lines. The ALMA beam size is illustrated by the ellipse in the lower left-hand corner.  The apertures used for the spatially resolved \irxb~analysis are shown by the white dashed lines.}
    \label{fig:IRX_Apertures}
\end{figure}

\begin{figure}
	\includegraphics[width=\columnwidth]{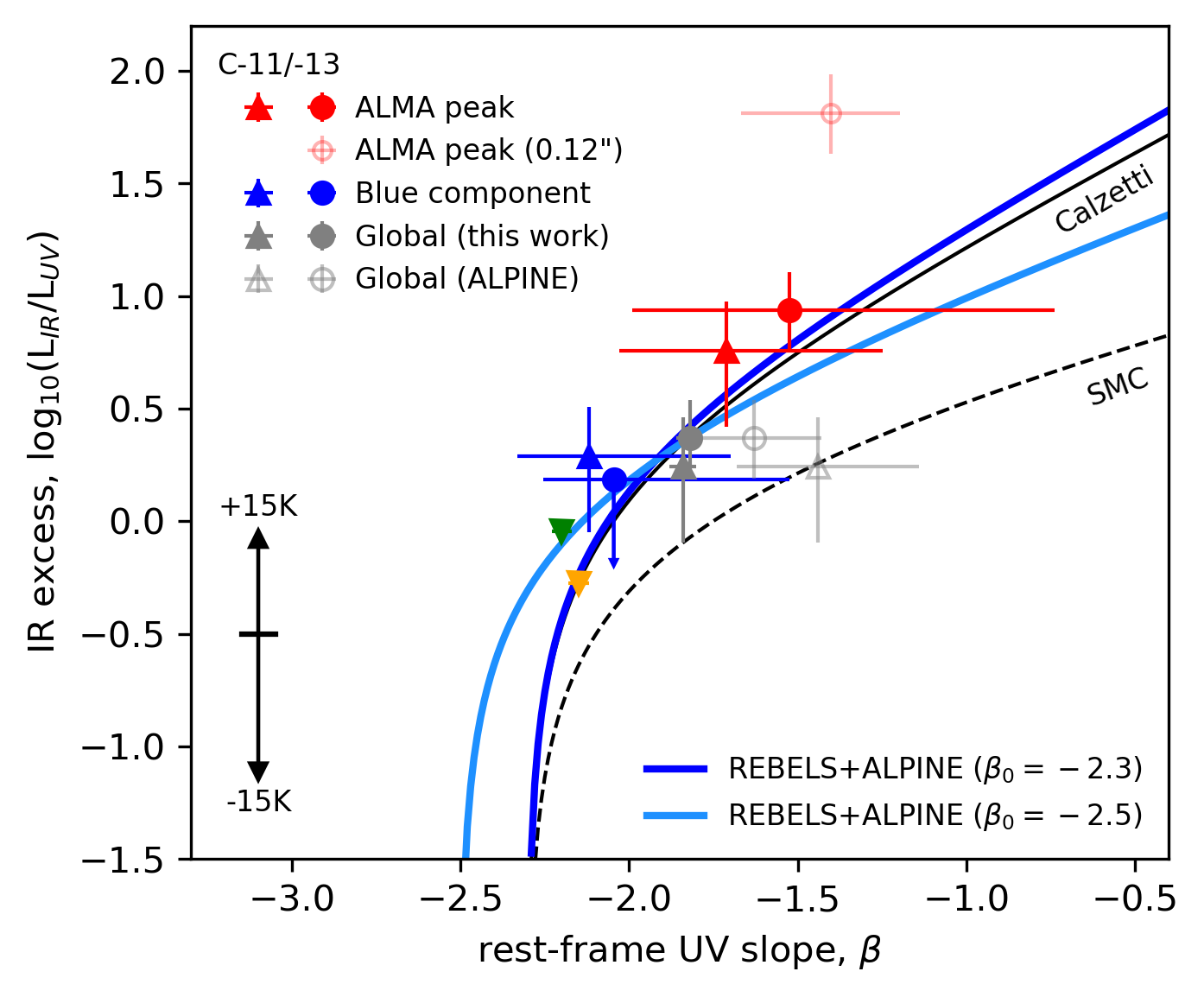}
    \caption{The {\irxb} relation points for CRISTAL-11 (triangles) and CRISTAL-13 (circles).  The values derived from the (ALPINE) resolved $\beta$ and {\Lir} from \citet{Mitsuhashi23} are shown in (open) grey.  
    Spatially resolved values using the apertures shown in Fig.~\ref{fig:IRX_Apertures} are shown in red (centred on the ALMA peak) and blue.  
    The upper limits for the CRISTAL-15 and CRISTAL-17 galaxies are shown as the yellow and green triangles and represent a $3\sigma$ limit on the \Lir.
    The expected relations for Calzetti-like dust attenuation and the SMC extinction law are shown as the solid and dashed black lines, respectively.  
    These assume an intrinsic $\beta-$slope of $\beta_{0} = -2.3$.
    The best-fitting {\irxb} relations to the REBELS and ALPINE galaxies from \citet{Bowler24} assuming an intrinsic UV-slope of $\beta_{0}=-2.3$ ($\beta_{0}=-2.5$) is shown by the solid dark (light) blue line. 
    The red open circle shows the value obtained if the ALMA continuum detection for CRISTAL-13 could be localised to within a $0.12$ arcsec diameter aperture.}
    \label{fig:Resolved_IRX}
\end{figure}

With the small sample of galaxies we study, that have both~\emph{JWST} PRIMER and ALMA-CRISTAL high-resolution observations, we can make some of the first measurements of the resolved \irxb~measurement within galaxies at $z \simeq 5$.
CRISTAL-11 and CRISTAL-13 are detected in the rest-frame 158$\mu$m dust continuum~ \citep{Mitsuhashi23}.  
These detections are shown by the contours on the {\av} map in Fig.~\ref{fig:c11}.  
The results of our resolved {\irxb} analysis using the apertures at the positions shown in Fig.~\ref{fig:IRX_Apertures} for these two galaxies are shown in Fig.~\ref{fig:Resolved_IRX}.
The apertures were chosen to enclose the flux from the distinct regions.
We measured a dust-associated {\irxb} value using a circular aperture centred on the peak of the ALMA dust continuum detection of diameters 0.30 arcsec for CRISTAL-11 and 0.48 arcsec for CRISTAL-13.  
A second value was obtained from an aperture encompassing the remaining emission.  
For each aperture, the resolved $\beta$ measurement was obtained by taking the mean of the values from the map shown in Fig.~\ref{fig:c11}.
The error bars correspond to the range of values in the aperture, and thus represent the spread in resolved $\beta$.
The fraction of the rest-frame UV flux enclosed by the aperture was calculated from the F090W image and was used to scale the {\muv} to obtain the aperture rest-frame UV luminosity.  
For the aperture centred on the ALMA dust emission peak, {\Lir} is taken from \cite{Mitsuhashi23} who exploit the same approach as used in the ALPINE analysis~\citep{Bethermin20} using a SED template derived from~\citet{Bethermin17} (equivalent to a modified blackbody with $T_{\rm d} \simeq 42\,{\rm K}$, $\beta_{\rm d} = 1.8$) integrated over $8$--$1000\,\mu{\rm m}$.
The dust continuum sizes measured in \citet{Mitsuhashi23} for CRISTAL-11 and CRISTAL-13 are $0.116 \pm 0.091$ arcsec and $0.092 \pm 0.070$ arcsec, respectively.  
Comparing these to the ALMA beam sizes we see that the dust continuua are barely resolved, suggesting the emission is coming from a compact region, smaller than the aperture sizes.
For the additional aperture, CRISTAL-11 has a $3\sigma$ detection in the dust continuum map on the south-east side of the galaxy, so this is presented as a detection in Fig.~\ref{fig:Resolved_IRX}.  
CRISTAL-13 has no detection associated with the western component (the `tail'), so this is presented as a (conservative) $3\sigma$ upper limit.  

We find that the rest-frame UV slope of all four sources are bluer than the previous values computed in~\citet{Faisst20} as shown in Table~\ref{tab:params}.
The~\citet{Faisst20} $\beta$ values were derived using the same method we exploit, fitting a power law to the best-fitting SED model to the photometry.
Thus, we find that the global IRX values are shifted bluewards in the \irxb~plane, now showing good agreement with the~\citet{Calzetti00} starburst relation.
The resolved measurements also appear to follow the predictions of this starburst dust law, although they do lie slightly above this relation potentially potentially due to further geometric effects on the scale of these clumps (e.g. see~\citealp{Popping17}).
These results are in agreement with the statistical measure of the \irxb~relations presented in~\citet{Bowler24}, where they found a Calzetti-like curve provided a good description of the REBELS and ALPINE samples.
One caveat in using the \irxb~relation is the unknown intrinsic rest-frame UV slope of the sample, which affects the position of the intercept of the relation with the $\beta$-axis.  
Thus, we show the fitted relations from \citet{Bowler24} for both $\beta_{0}=-2.3$ and $\beta_{0}=-2.5$, since these values denote the range of values inferred from the SED fits to the REBELS and ALPINE sources. 
Another major caveat in this analysis is the uncertain dust SED at high redshift, which as discussed in~\citet{Bowler24} and~\citet{Mitsuhashi23}, can lead to systematic offsets in IRX of up to 0.4 dex with a 10K difference in the assumed $T_{\rm d}$ (keeping the emissivity index, $\beta_{\rm d}$, fixed).
Individual dust temperature measurements are required to understand the exact \irxb~relation, for example recent work on HZ10 by~\citet{Villanueva24}, however our results demonstrate that an \irxb~relationship similar to the local starburst relation of~\citet{Calzetti00} appears to hold at a resolved level at $z \simeq 5$.

\subsection{The role of dust in shaping the rest-UV morphology}

Given the excellent spatial resolution of the ALMA-CRISTAL data, we can make further tentative investigations into the location of the dust within CRISTAL-11 and CRISTAL-13 in comparison to the rest-frame UV light.
The other two sources we study were not detected in the ALMA-CRISTAL survey in the dust continuum~\citep{Mitsuhashi23}.
From the ALPINE and REBELS surveys it has been shown that the dust obscured SFR (or obscured fraction) drops rapidly around \lmstar $\simeq 9.5.$~\citep{Fudamoto20, Algera23, Bowler24}, and hence the fact that CRISTAL-15 and CRISTAL-17 are undetected (especially given the low S/N of the other detections) is unsurprising.
Furthermore, these sources are bluer than CRISTAL-11 and CRISTAL-13, with CRISTAL-17 also having a significantly lower SFR and higher redshift.
Despite this, we see regions of comparable \av~across the sources, and deeper ALMA data may yet reveal dust continuum detections at these positions.

In Fig.~\ref{fig:IRX_Apertures} we show the CRISTAL contours in comparison to the bluest (and hence highest resolution)~\emph{JWST} NIRCam band, F090W.
This comparison shows that the CRISTAL-11 peak dust detection lies offset by approximately 0.6 arcsec from the rest-frame UV peak in this source, the same magnitude of offset that is found for 30 percent of the ALPINE sample in the statistical analysis of~\citet{Killi24}.
We confirm that this rest-frame UV to FIR spatial offset is due to a young star-forming clump $\simeq 4\,{\rm kpc}$ away from the stellar mass centroid, which is also the dustiest region of the galaxy (as revealed by the SED fitting derived \av~and the direct detection of the dust continuum).

We find that in CRISTAL-11 the offset between the rest-FIR and UV emission can be naturally explained by differences in amount of dust attenuation (there is also a difference in age between these two clumps but it is small, only $\sim 50\,{\rm Myr}$).
In CRISTAL-13, the dust continuum detection is offset from the bluest regions of the galaxy (in the Western tail), and instead follows the \mstar~map closely.
If we closely inspect the map of CRISTAL-13 shown in Fig.~\ref{fig:IRX_Apertures}, we find a small offset between the ALMA dust continuum detection for CRISTAL-13 and the rest-frame UV bright NE clump of the order of 0.15 arcsec.
If the dust continuum detection is in fact originating from the region offset from the NE clump, then this could point to dust attenuation leading to the gap in the observed rest-frame UV emission from this source.
This dust-star morphology is similar to that seen in the `Cosmic Grapes' sources, where~\citet{GimenezArteaga24} find that the observed clumps appear to have similar ages but differ in their derived \av, indicating that differing dust coverage does shape the observed rest-frame UV morphology (although see~\citealt{Fujimoto24, Zanella24} who suggest a reduced importance of dust). 
These results are in close agreement with the recent simulation work of~\citet{Ocvirk24} and~\citet{Nakazato24} (see Section~\ref{sect:sims}), who found that dust attenuation can shape the observed rest-frame UV morphology, and lead to offsets between the UV and FIR peak.

\begin{table*}
   \caption{The ALMA-CRISTAL derived far-infrared properties of our galaxy sample.
   In Column 1 we show the galaxy ID, followed by the component of interest in Column 2 and the absolute UV magnitude in Column 3.
   The rest frame $158$ $\mu$m flux, with the SNR shown in brackets, is shown in Column 4, followed by the infrared luminosity in Column 5.
   The FIR measurements are reproduced from \citet{Mitsuhashi23}, as is the FIR size shown in the final Column 10.
   The global and aperture (see Fig.~\ref{fig:IRX_Apertures}) {\irx} values are given for the galaxies with ALMA dust detections (CRISTAL-11/-13) in Column 6, with the rest-frame UV-slope for that region presented in Column 7. 
   The mean UV-slope for each sub-component was obtained from the $\beta$ maps shown in Fig.~\ref{fig:c11}, and the errors correspond to the range of values within each aperture.
   Columns 8 and 9 correspond to the non-parametric sizes obtained for the full galaxy from~\citet{Varadaraj24}.}
    \centering
    \begin{tabular}{ccccccccccccc}
    \hline
        ID & Component & \muv  & $S_{\nu}$  & $\log_{10}$(\Lir/\Lsun) & \irx & $\beta$ & $R_{\rm F115W}$ & $R_{\rm F444W}$  & $R_{\rm FIR}$\\
         & & /mag & /$\mu$Jy & & & & /{\rm kpc} & /{\rm kpc} & /{\rm kpc} \\
        \hline
        CRISTAL-11 & Global & -21.35 & $137 \pm 55 (4.6)$ & $11.11\substack{+0.19 \\ -0.33}$  & $0.24\substack{+0.22 \\ -0.34}$ & $-1.84\substack{+0.04 \\ -0.04}$  & $0.91 \pm 0.16$ & $0.90 \pm 0.11$ & $0.77 \pm 0.60$ \\[1ex]
        & N & -20.07 & -- & $11.11\substack{+0.19 \\ -0.33}$ & $0.76\substack{+0.22 \\ -0.34}$ & $-1.7\substack{+0.5 \\ -0.3}$ & & & \\[1ex]
        & S & -20.77 & -- & $10.92\substack{+0.19 \\ -0.33}$ & $0.29\substack{+0.22 \\ -0.34}$ & $-2.1\substack{+0.4 \\ -0.2}$ & & & \\[1ex]
        CRISTAL-13 & Global & -21.34 & $137 \pm 39$ (6.6) & $11.23\substack{+0.12 \\ -0.16}$  & $0.37\substack{+0.17 \\ -0.18}$ & $-1.82\substack{+0.04 \\ -0.04}$ & $1.33 \pm 0.12$ & $1.62 \pm 0.07$ & $0.60 \pm 0.46$ \\[1ex]
        & NE & -19.92 & -- & $11.23\substack{+0.12 \\ -0.16}$ & $0.94\substack{+0.17 \\ -0.18}$ & $-1.5\substack{+0.8 \\ -0.5}$ & \\[1ex]
        & WT & -20.94 & -- & $< 10.89$ & $< 0.18$ & $-2.0\substack{+0.5 \\ -0.2}$  & & & \\[1ex]
        CRISTAL-15 & Global  & -21.90 & $< 104$ & $\leq 10.81$  & $\leq -0.27$ & $-2.15^{+0.03}_{-0.03}$ & $0.92 \pm 0.14$ & $1.10 \pm 0.09$ & --\\[1ex]
        CRISTAL-17 & Global & -21.37 & $< 51 $ & $\leq 10.83$  & $\leq -0.04$ & $-2.20^{+0.03}_{-0.03}$ & $1.26 \pm 0.11$ & $1.10 \pm 0.10$ & --\\
        \hline
    \end{tabular}
    \label{tab:fir}
\end{table*}

Given the beam size and SNR of the ALMA detection for the CRISTAL-13 source, the error on the position of the peak can be approximated as $\simeq {\rm FWHM}/{\rm SNR} = 80\,{\rm milliarcsec}$.  
From~\citet{Mitsuhashi23} this ALMA detection is compact, with a measured size of $0.092 \pm 0.070$ arcsec.
Therefore, we can use a smaller aperture of diameter 0.12 arcsec to determine the \Luv~and $\beta$, centred on the peak of the ALMA detection.
We correct \Luv~for the effect of aperture loss in the F090W band due to the PSF using a factor of $0.69$.
In this smaller aperture, we unsurprisingly recover a redder rest-frame UV slope and a higher IRX as shown with the open circle in Fig.~\ref{fig:Resolved_IRX}.
This point lies above the canonical \irxb~relations, indicating a highly obscured region where the observed rest-frame UV colour is not adequately describing the obscured SFR fraction, for example in the case of a dust screen with holes.
A similar offset is also found for regions of the luminous $z = 5.67$ galaxy HZ10 as presented in~\citet{Villanueva24}, where they are able to derive resolved dust temperature estimates thus reducing the uncertainty on the IRX.
We caution that the resolution of the ALMA data is insufficient to reliably ascertain if these features are spatially offset, however this analysis demonstrates that an inhomogeneous dust distribution within CRISTAL-11 and CRISTAL-13 can explain the observed rest-frame UV morphology, with differences in age being a secondary effect.
This hypothesis is also supported by the observed negative correlation between the rest-frame UV to FIR offsets and the obscured fraction of star formation found for the full CRISTAL sample~\citep{Mitsuhashi23}.

\section{A comparison to the FirstLight simulation}\label{sect:sims}

\begin{figure*}
    \centering
    \includegraphics[width=\textwidth]{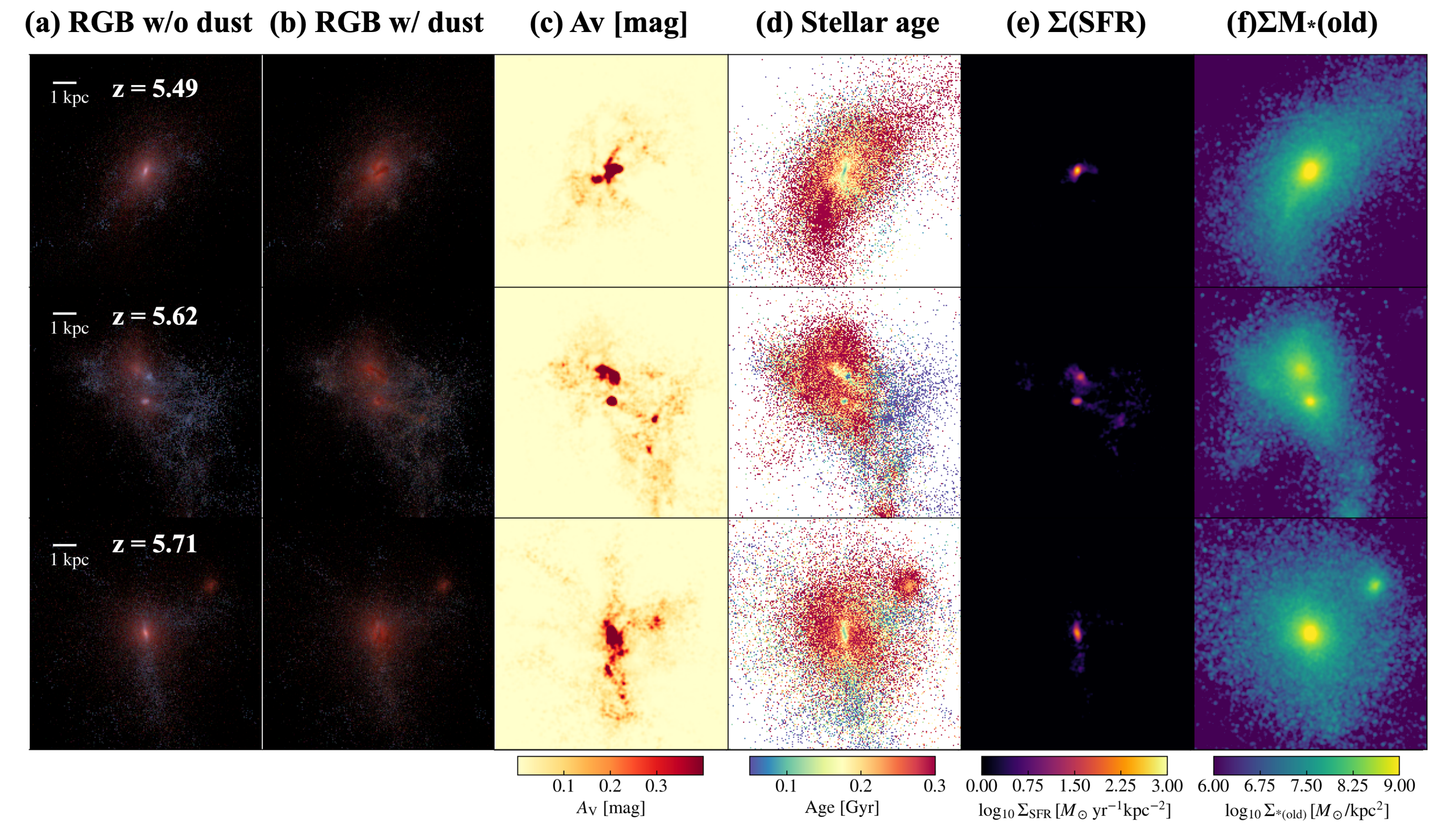}
    \caption{Three $z \simeq 5.5$ galaxies found in the FirstLight simulation suite~\citep{Ceverino17} selected from the~\citet{Nakazato24} sample.
    The galaxies have stellar masses of \lmstar$ = 9.3$--$9.6$ and SFRs of $\simeq 20\,$\Msun/yr, closely comparable to the CRISTAL galaxies we study in this work.
    Each stamp has dimensions of $10 \times 10 {\bf kpc}$.
    The first column (a) shows three-color (NIRCam F115W, F200W, and F356W for RGB) mock images without dust attenuation. 
    The second column shows the same RGB images as in column (a), however here dust attenuation is included. 
    Column (c) shows the V-band dust attenuation (\av), followed by column (d) showing the mass-weighted stellar age.
    To aid in comparison, the colour bar for these parameters matches that used CRISTAL-17 in Fig.~\ref{fig:c17}.
    Finally columns (e) and (f) show the SFR density and stellar mass distribution (of stars older than 100 Myrs) respectively.}
    \label{fig:sim}
\end{figure*}

In this section we compare our results to the zoom-in simulation, FirstLight~\citep{Ceverino17}.
\citet{Nakazato24} use the FirstLight simulations to provide a spatially resolved analysis of high-redshift galaxies.
In particular, they simulated galaxies of comparable stellar masses to the objects we study in this work allowing for a close comparison with our~\emph{JWST} results.
\citet{Nakazato24} identify 100 parsec scale clumps within their simulated galaxies, similar to the morphology and sizes of clumps that we identify in the CRISTAL galaxies.
They found that between 10--20 percent of galaxies with \mstar $= 0.1-5 \times 10^{10}$\Msun~were observed in a phase dominated by major-merger activity, leading to the observed clump formation, in the redshift range of their work ($z \sim 5$ to $z \sim 9.5$).
While this fraction is lower than we recover, the difference could be due to the definition of clumps in~\citet{Nakazato24} who used a threshold in star-formation surface density on the native resolution simulations.
This merger driven clump formation within high-redshift galaxies has also been found in other simulations, such as~\citet{Ocvirk24},~\citet{Ma19} and~\citet{Pallottini22}.

Interestingly, the zoom in cosmological simulations are able to recover galaxies with similar extent and clumpy morphology to the CRISTAL galaxies.
Figure~\ref{fig:sim} shows three galaxies from the FirstLight simulation at $z \simeq 5.5$, with stellar masses and SFRs that closely match those in our sample (\lmstar $= 9.3$--$9.6$; SFR $=19$--$22$\Msun/yr). 
The SFRs shown are calculated with stars younger than 10 Myrs.
The broad morphology of old (> 100 Myr) and young (< 10 Myr) stars are consistent in the simulated sources, with both the old and young stars showing clumps separated by $1$-$4$ kpc within an extended structure.
This is very similar to what we observe for the CRISTAL galaxies and hence we can draw parallels with the formation scenario of the simulated galaxies and our observed galaxies.
~\citet{Nakazato24} identify two broad populations of clumps, with the majority being formed as a result of major mergers in the simulation.
The first class of clumps were found to be older (ages $> 50\, {\rm Myr}$) and represent proto-bulges, with the second class being younger clumps that have high sSFRs of $5 \times 10^{-8}{\rm yr}^{-1}$ and are formed within the tidal tails of the interacting systems.
We recover both older regions of several tens of Myr, in addition to younger regions that show comparable sSFRs to the simulations (e.g. Fig.~\ref{fig:c11}).
In addition, the simulated images show the importance of dust emission on the observed morphology, with the compact cores in the simulated galaxies becoming more diffuse when dust is applied.
Our analysis of the ALMA-CRISTAL dust continuum data in Section~\ref{sect:disdust} shows how dust is likely influencing the observed morphology of the most massive galaxies we study here, by obscuring certain regions that appear fainter and redder in the rest-frame UV than other parts of the galaxy.

Overall, the similarity of the CRISTAL galaxies we study to the sources found in the cosmological simulation presented in~\citet{Nakazato24} allows us to tentatively conclude that the multi-wavelength morphology of our sample could have been impacted by major mergers.
This conclusion is supported by the initial morpho-kinematic classes of these sources defined in~\citet{LeFevre20} from the ALPINE survey, which indicated dispersion dominated sources.
It is now supported by the first kinematic results from the CRISTAL survey, which also find evidence for major mergers in the \cii~distribution~\citep{Posses24, Solimano24}.
A detailed analysis of the full \cii~kinematics from the CRISTAL program (Lee et al. in prep) will shed more light onto the formation mechanism of \lmstar$\simeq 9.5$ galaxies in the early Universe.
A caveat of this conclusion is that in the FirstLight simulation only 10--20 percent of sources show the clumpy morphology that we see in all four of the CRISTAL galaxies in this study.
One reason for this could be the different clump selection procedures between the simulation and observational analyses, and further investigation is required to match the population statistics and small scale morphological measurements in models and data (e.g. see the detailed size-luminosity comparison in~\citealp{Varadaraj24}).

\section{Conclusions}\label{sect:conclusions}
In this work we perform a pixel-by-pixel analysis of four normal galaxies (with \lmstar $ \simeq 9.5$) at $z = 4.5$--$5.6$ using the multi-band NIRCam imaging from the~\emph{JWST} PRIMER program.
The four sources were previously spectroscopically confirmed as part of the ALPINE program, and have deep, high-spatial resolution, dust continuum observations from the ALMA-CRISTAL large program.
Using an SED fitting analysis with a simple constant SFH model we study the resolved and integrated properties and investigate the potential biases present in fitting to unresolved photometry of galaxies of this mass.
Our main conclusions are as follows:

\begin{itemize}
    \item Compared to the previously available~\emph{HST} optical and NIR data, we find that all of the sources break into further kpc-scale clumps and show at least 2 (and up to 8) separate components in the high-resolution~\emph{JWST} data.
    We find comparable morphologies in the rest-frame UV and optical, with no clear evidence for underlying older disk structures.
    The non-parametric sizes of the sources are $\sim 1\,{\rm kpc}$ as expected from the size-mass relation, however the full extent of the galaxies is $\simeq 5$--$10\,{\rm kpc}$ taking into account the widely separated clumps.
  
    \item In contrast to some studies of lower mass galaxies, including the strongly lensed Cosmic Grapes source, we do not find extremely young ages (or high rest-frame optical emission lines) or other evidence for strong outshining of the observed global emission by the youngest regions.
    We find that the stellar masses derived from the resolved pixel-by-pixel analysis only slightly higher ($0.1$--$0.3\,{\rm dex}$) than those determined from fitting to the global integrated photometry.
    Our results agree with those found in star-forming galaxies of similar mass at $z \simeq 2$~\citep{Shen24} and $z \gtrsim 4$~\citep{PerezGonzalez23}.
    The result holds for both an assumed constant and delayed$-\tau$ star-formation history.
    
    \item In comparison to the previously published best-fitting properties of these galaxies, our~\emph{JWST} analysis recovers younger ages ($\simeq 100\,{\rm Myr}$) and bluer rest-frame UV slopes ($\beta \simeq -2.1$).
    The~\chone~photometry from the COSMOS2020 catalogue was found to be overestimated by a factor of $\gtrsim 2$ for CRISTAL-17, likely due to confusion in the low resolution~\emph{Spitzer}/IRAC data.
    We further recover slightly weaker \halpha~emission than the results obtained in the ALPINE catalogues (\ew$\simeq 400$\AA).
    \item For two of the sources we compare the physical parameter maps to the resolved dust continuum measurements from the ALMA-CRISTAL program.
    In both cases we see a strong correlation between the direct detection of dust and the dust attenuation derived from the~\emph{JWST} photometry in the rest-frame UV and optical, as parametrised by the rest-frame UV slope and \av. 
    The peak of the dust continuum also correlates with the centre of mass of the galaxies, as predicted by the recent simulation work of~\citet{Ocvirk24}.
    \item We compute the resolved \irxb~relation for CRISTAL-11 and CRISTAL-13 to decouple the effect of geometrically separate regions on the correlation.
    We find bluer rest-frame UV slopes than that determined in the ALPINE program (from COSMOS2015 photometry).
    With the assumed dust temperature and $\beta_{\rm d}$ parameters from~\citet{Mitsuhashi23}, we recover a correlation between \irx~and colour that is in good agreement with the local starburst~\citet{Calzetti00} relation.
    As expected, the position of the ALMA-CRISTAL dust detection is co-spatially with the redder regions of the galaxies, suggesting that dust attenuation is shaping the observed rest-frame UV morphology.
    
    \item We extracted a sample of three galaxies at $z \sim 5.5$ from the FirstLight simulation with the same stellar masses and SFRs as the CRISTAL sources to investigate the formation mechanism of such galaxies.
    Interestingly, we recover similar age and colour gradients between the simulations and observations, as well as comparable physical extents ($> 1\,{\rm kpc}$, with several sources showing clumpy morphologies in the simulations).
    Major mergers are seen to be important in the formation of clumps in FirstLight, and hence we tentatively conclude that mergers could be important in the formation of clumpy massive galaxies at early times.

\end{itemize}
These results demonstrate the power of combining~\emph{JWST} resolved photometry with high-spatial resolution ALMA imaging to understand galaxies at high redshift.
Additional insight into the CRISTAL galaxies will be obtained with upcoming NIRSpec IFU spectra (PI: Faisst, PID: 3045 and PI: Aravena, PID: 5974), which will allow detailed measurements of the stellar ages and ioniziation conditions from the rest-frame optical emission line ratios.

\section*{Acknowledgements}
RB acknowledges support from an STFC Ernest Rutherford Fellowship [grant number ST/T003596/1].
CTD acknowledges the support of the Science and Technology Facilities Council. 
DC thankfully acknowledges the computer resources at MareNostrum and the technical support provided by the Barcelona Supercomputing Center (RES-AECT-2020-3-900 0019).
DC is supported by the Ministerio de Ciencia, Innovacion y Universidades (MICIU/FEDER) under research grant PID2021-122603NB-C21.
DJM acknowledges the support of the Science and Technology Facilities Council. 
RH-C thanks the Max Planck Society for support under the Partner Group project "The Baryon Cycle in Galaxies" between the Max Planck for Extraterrestrial Physics and the Universidad de Concepción. 
RH-C, MA, RJA, MK and MS acknowledge financial support from ANID BASAL project FB210003. 
RJA was supported by FONDECYT grant number 1231718. 
MS was financially supported by Becas-ANID scholarship \#21221511. 
JSD acknowledges the support of the Royal Society through a Royal Society Research Professorship. 
VV acknowledges support from the ALMA-ANID Postdoctoral Fellowship under the award ASTRO21-0062. 
RI is supported by Grants-in-Aid for Japan Society for the Promotion of Science (JSPS) Fellows (KAKENHI Grant Number 23KJ1006). 
AF acknowledges support from the ERC Advanced Grant INTERSTELLAR H2020/740120. 
PGP-G acknowledges support from grant PID2022-139567NB-I00 funded by Spanish Ministerio de Ciencia, Innovaci\'on y Universidades MCIU/AEI/10.13039/501100011033,
FEDER {\it Una manera de hacer Europa}.
MR acknowledges the support from the project PID2020-114414GB-100, financed by MCIN/AEI/10.13039/501100011033.
YN acknowledges funding from JSPS KAKENHI Grant Number 23KJ0728.

\section*{Data Availability}
The PRIMER and CRISTAL data are available in the public domain.



\bibliographystyle{mnras}
\bibliography{bibtex_parsedcristal} 




\newpage
\appendix
\section{Comparison to Li+2024}\label{appendix:li}
In Table~\ref{tab:li} we present the resolved and integrated \mstar~and SFRs from the fitting analysis of Li et al. (2024).
In Li et al. (2024) they performed SED fitting using {\tt MAGPHYS}, of a larger sample of CRISTAL galaxies using the full rest-frame UV to FIR data including that from~\emph{JWST} and the ALMA-CRISTAL program.
They matched the spatial resolution of the~\emph{JWST} imaging to that of the available ALMA data, and hence this provides a complementary fitting procedure (at slightly lower spatial resolution) than our~\emph{JWST}-only analysis.
Reassuringly we find excellent agreement between the stellar masses, with the resolved values being within $0.1\,{\rm dex}$.
While the integrated value agrees well for CRISTAL-11, we find a slightly lower (by 0.25\,{\rm dex}) value for CRISTAL-13 although this is only $2.5\sigma$ discrepant.
The SFRs are all consistent within the errors.
The resolved SFRs are slightly lower in our analysis, which could be due to not including the ALMA FIR data within our SED fitting.
The difference is only $\sim 5\,$\Msun$/{\rm yr}$ however, demonstrating that in these sources the fitting of~\emph{JWST} only photometry is able to recover over 80 percent of the total SFR (as defined as that from the full rest-frame UV to FIR fitting).
This result is consistent with our \irxb~analysis, which shows that first, the galaxies are relatively blue, suggestive of a moderate impact of dust obscuration on the observed rest-frame UV light.
And second, that the~\citet{Calzetti00} law provides a good description of the~\irxb~relationship.
We fit assuming this dust law in our {\tt BAGPIPES} fitting, and this appears to account for the majority of obscured SFR when fitting to the rest-frame UV and optical light alone.

\begin{table}
    \centering
     \caption{The resolved and integrated stellar masses and SFRs from the complementary analysis by Li et al. (2024).}
    \begin{tabular}{llcc}
\hline
& & CRISTAL-11 &  CRISTAL-13 \\
\hline
\lmstar & Resolved & $9.75^{+0.10}_{-0.13}$ & $ 9.72^{+0.09}_{-0.11} $\\[1ex]
& Integrated & $9.68^{+0.11}_{-0.08}$ & $9.75^{+0.07}_{-0.10}$ \\
\hline
SFR [\Msun/${\rm yr}$] & Resolved & $30.0 \pm 10.1$ & $30.1 \pm 7.8$ \\[1ex]
&  Integrated & $30.0^{+11.4}_{-8.0}$ & $ 23.8^{+8.3}_{-5.1}$ \\
 \hline
    \end{tabular}
   
    \label{tab:li}
\end{table}

\section{Delayed exponential SFH}\label{appendix:sfh}
Here we present the corresponding resolved SED fitting parameter and integrated fitting analysis plots when assuming a delayed-$\tau$ SFH.

\begin{figure*}
    \includegraphics[width=0.9\textwidth]{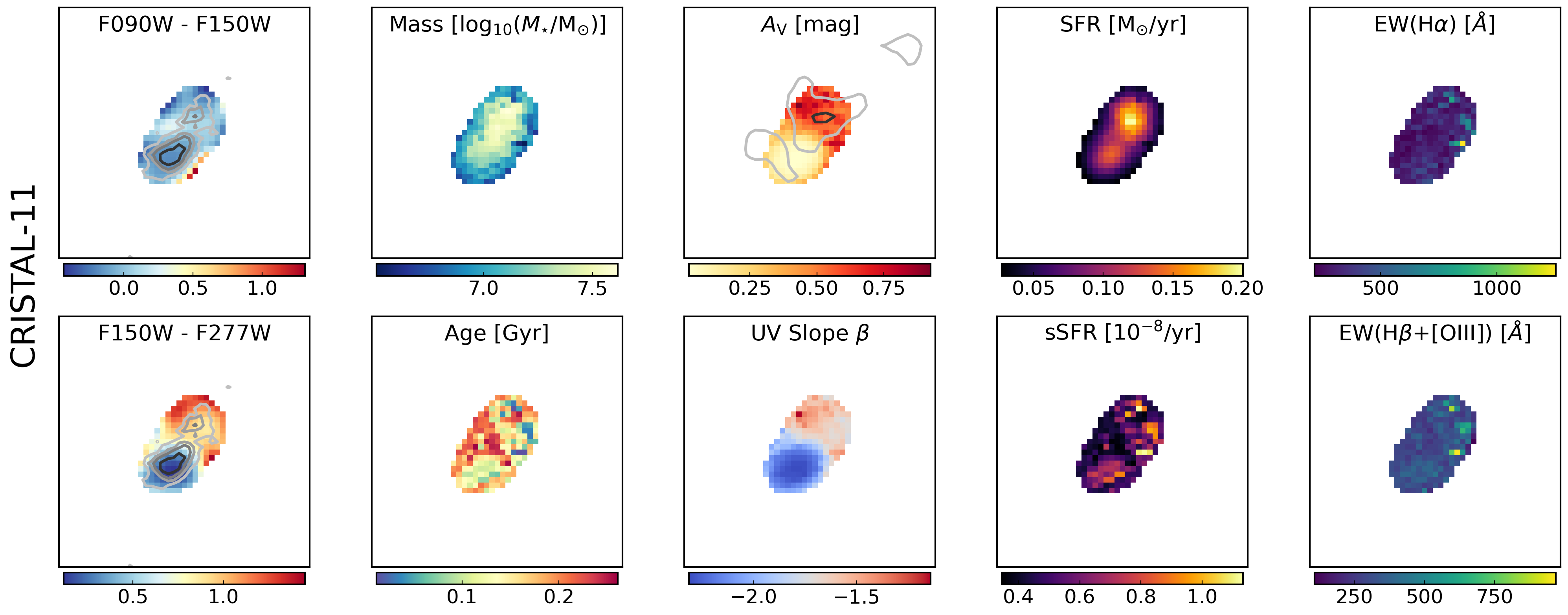}
    
    \vspace{0.5cm}
    
     \includegraphics[width=0.9\textwidth]{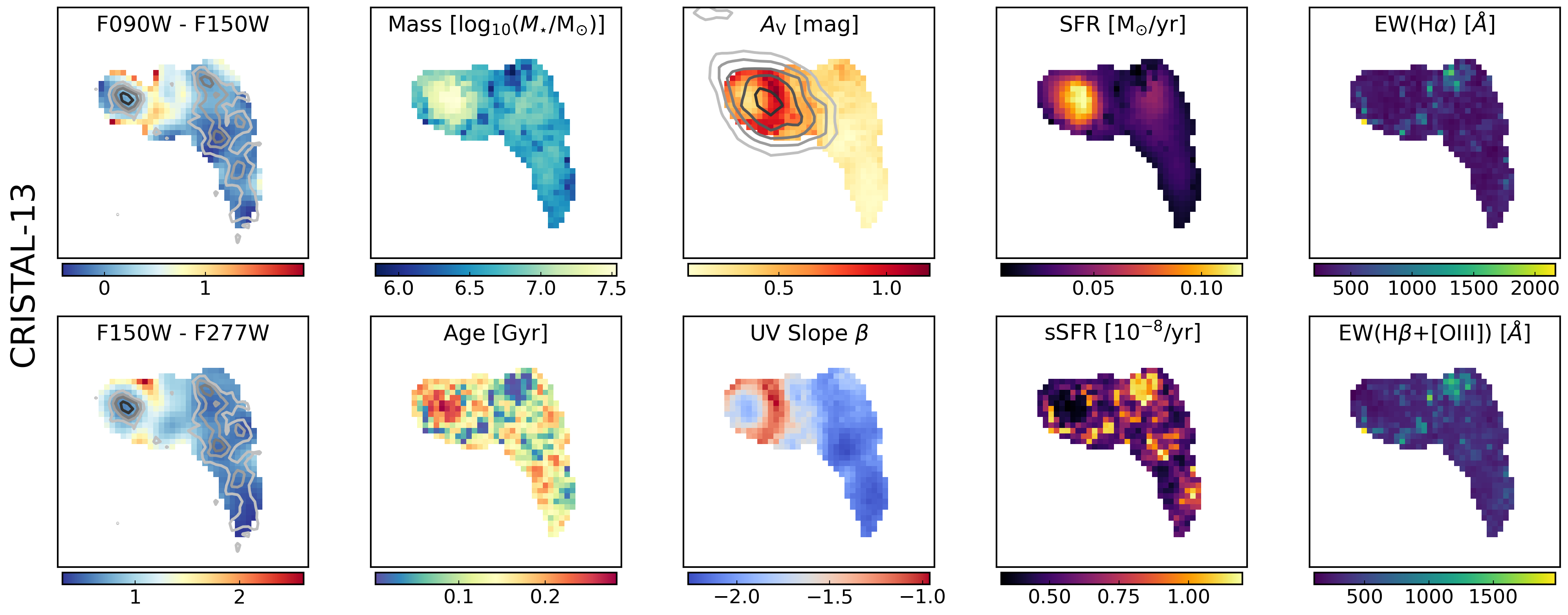}
     
     \vspace{0.5cm}
     
    \includegraphics[width=0.9\textwidth]{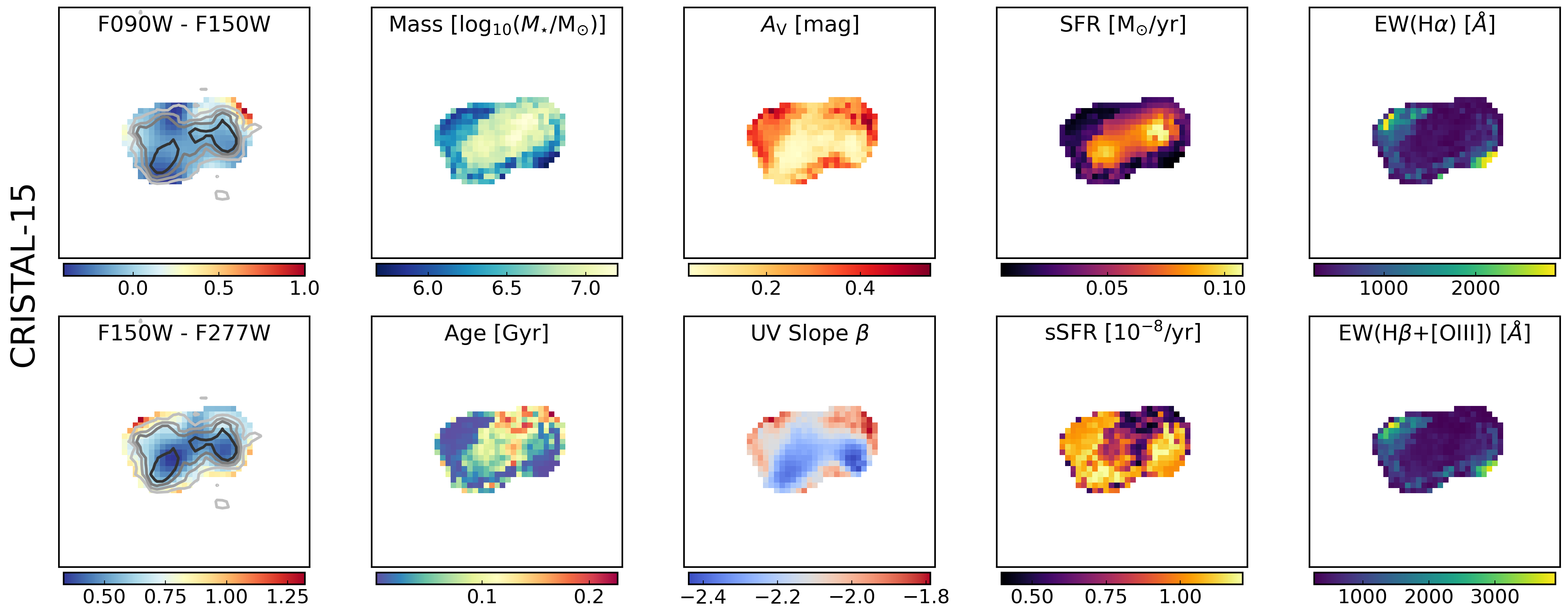}
\caption{The $z \simeq 4.5$ galaxies CRISTAL-11, CRISTAL-13 and CRISTAL-15 fitted with a delayed-$\tau$ assumed SFH.  
The details of this figure are described in the caption to Fig.~\ref{fig:c11}.}
    \label{fig:delayed-tau-c11}
\end{figure*}

\begin{figure*}
    \includegraphics[width=0.9\textwidth]{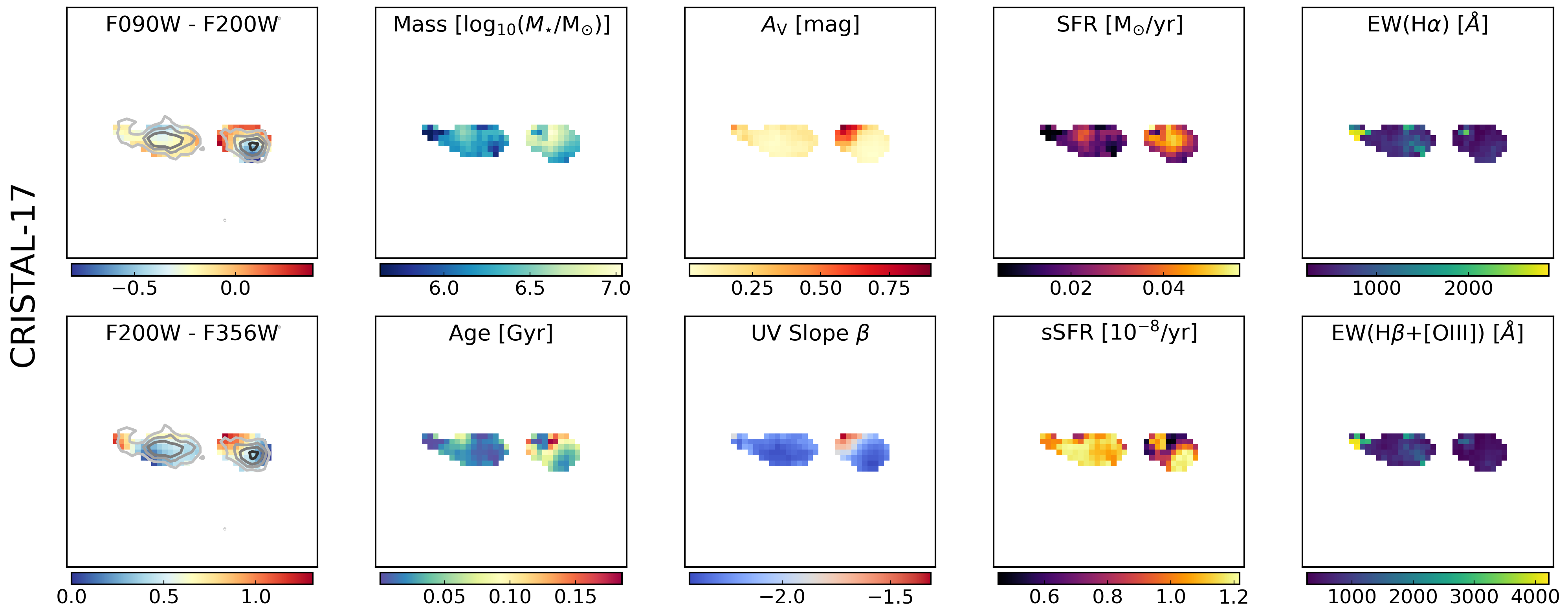}
\caption{The $z \simeq 5.5$ galaxy CRISTAL-17 fitted with a delayed-$\tau$ assumed SFH. 
The details in this figure are described in the caption to Fig.~\ref{fig:c17}.}
    \label{fig:delayed-tau-c17}
\end{figure*}

\begin{figure*}
    \centering
    \includegraphics[width = 0.49\linewidth]{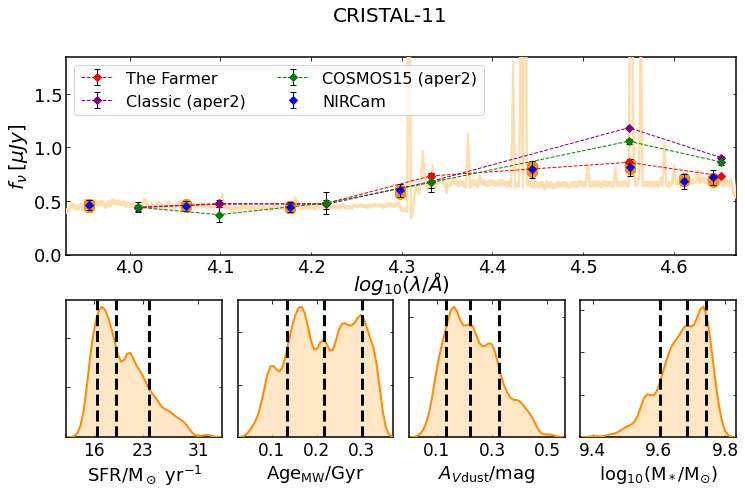}
       \includegraphics[width = 0.49\linewidth]{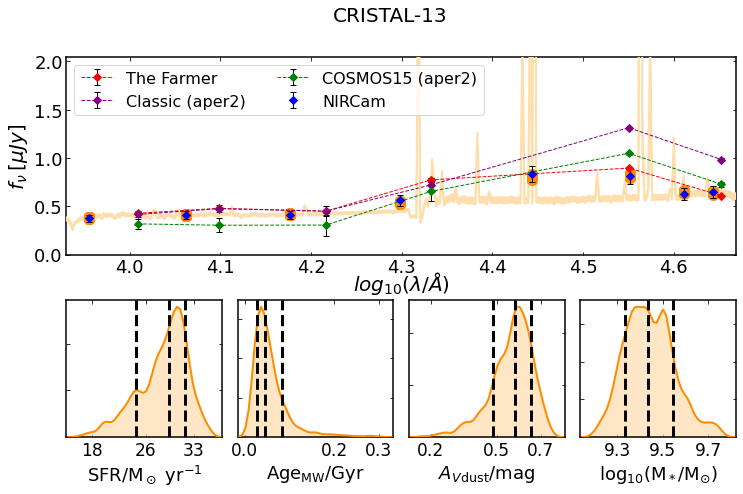}\\
        \includegraphics[width = 0.49\linewidth]{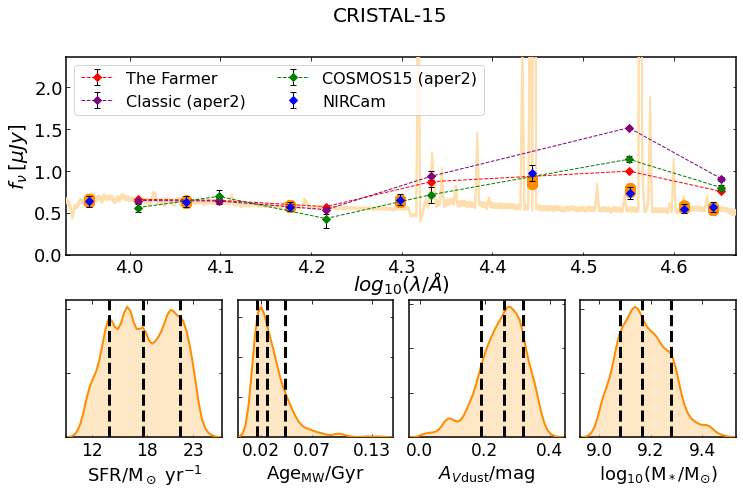}
       \includegraphics[width = 0.49\linewidth]{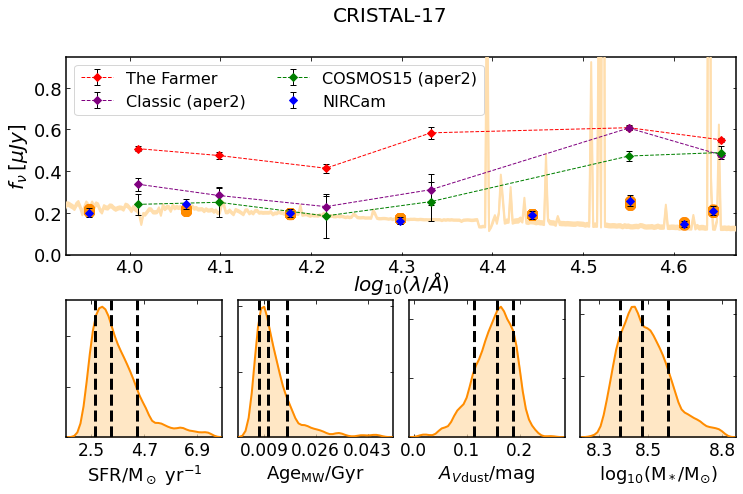}\\
    \caption{The results of the integrated SED fitting with a delayed-$\tau$ SFH.
    The figure properties are described in the caption to Fig.~\ref{fig:integrated-sed}.}
    \label{fig:delayed_integrated SED}
\end{figure*}

\section{Corner plots for pixel fits}\label{appendix:pixel}

Here we present corner plots as derived from {\sc BAGPIPES} for individual pixels in our resolved fitting.
The relevant pixels, chosen to represent regions of the galaxy with different derived properties, are highlighted in Fig.~\ref{fig:c11} and Fig.~\ref{fig:c17}.
We recover well known degeneracies in our fitting, in particular the degeneracy between age and stellar mass, and to a lesser degree dust \av~and stellar mass.
The coverage of strong rest-frame optical emission lines reduces the effect of these degeneracies, as the strength of the \halpha~and~\hboii~lines we infer from our photometry constrains the age in the individual pixel (and integrated) fits.
In some cases we hit the edge of parameter space, in particular for stellar age.  
Demonstrating the likely presence of very young stellar populations within the sources, as expected for star-forming regions in UV-bright LBGs.

\begin{figure*}
  \centering
    \includegraphics[width = 0.49\linewidth]{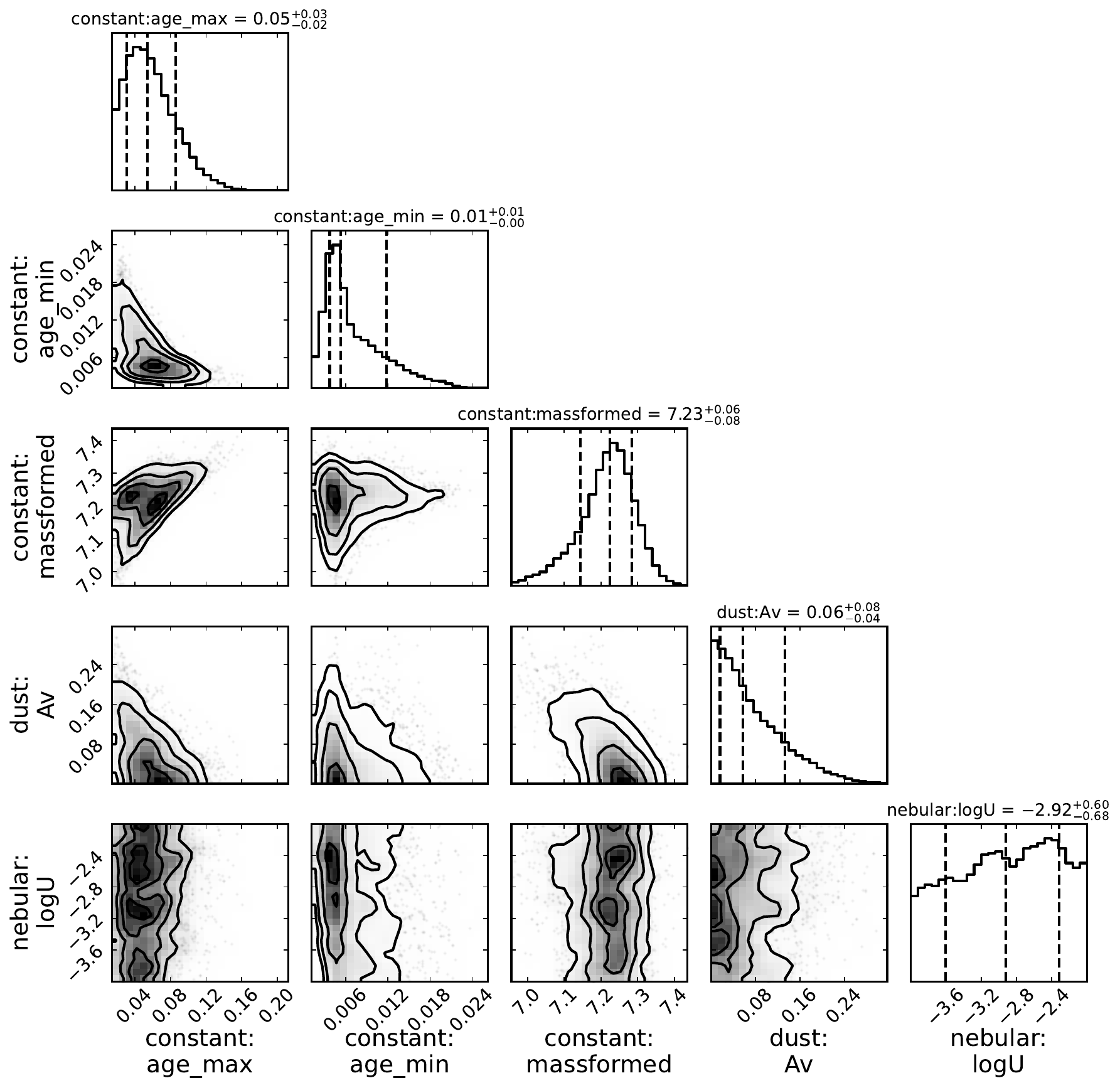}
       \includegraphics[width = 0.49\linewidth]{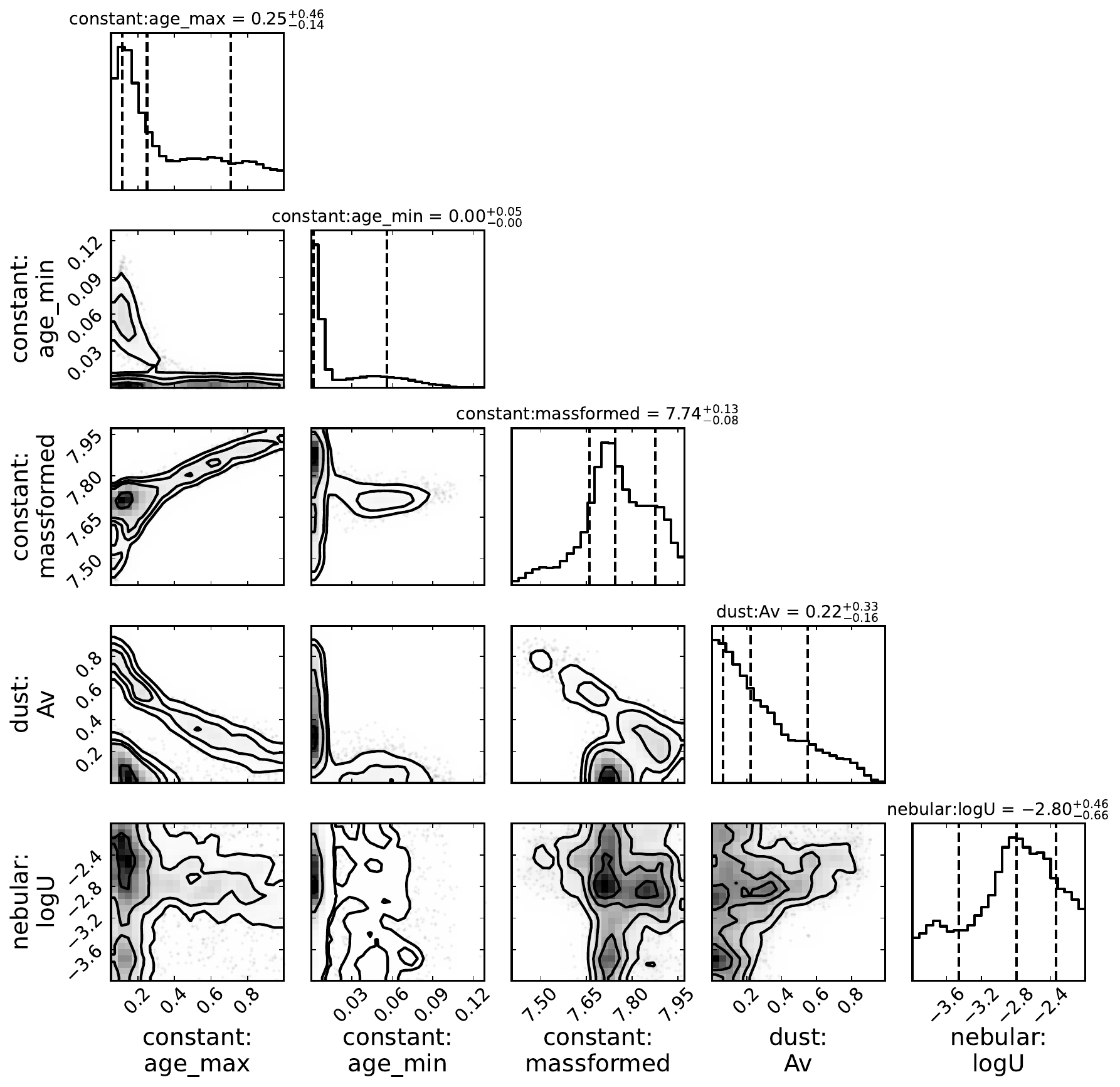}\\
         \includegraphics[width = 0.49\linewidth]{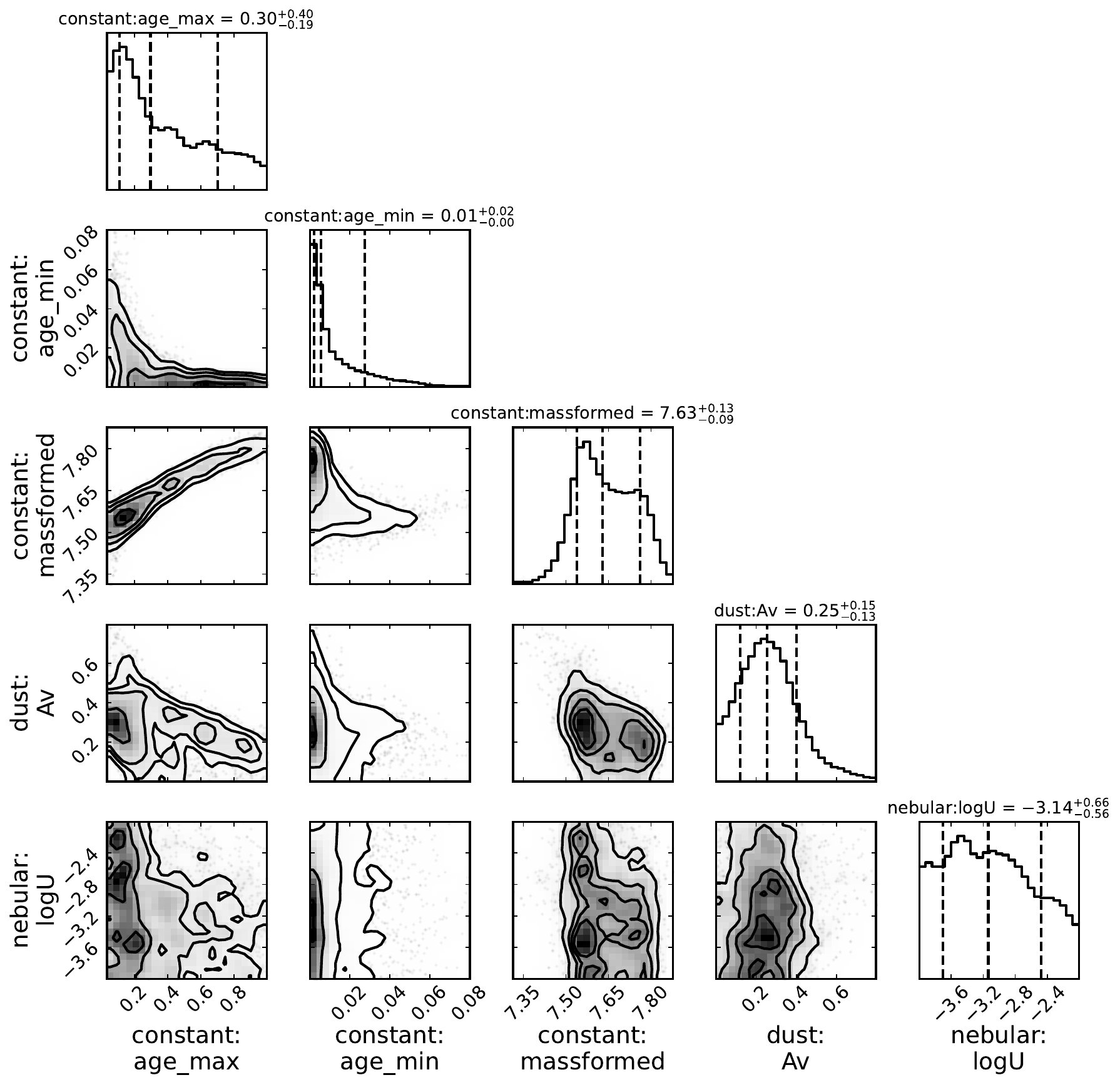}
       \includegraphics[width = 0.49\linewidth]{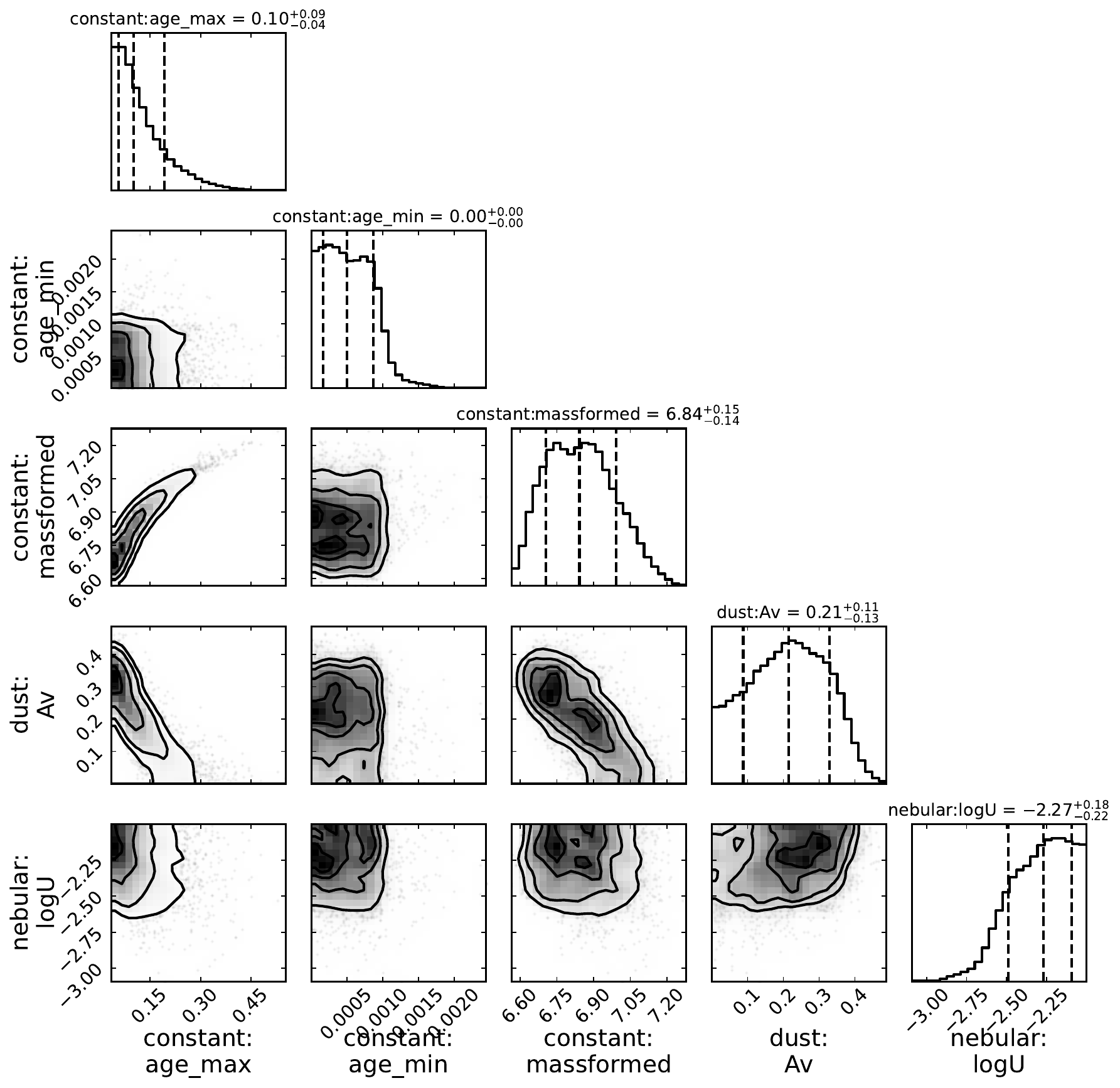}\\
    \caption{Corner plots derived from {\sc BAGPIPES} for CRISTAL-11 (top row) and CRISTAL-13 (bottom row) for the SED fitting of individual pixels. A CSFH was assumed.
    The relevant pixels are highlighted in Fig.~\ref{fig:c11}, with the left column corresponding to the red highlighted pixels and the right column corresponding to the orange highlighted pixels.
    }
    \label{fig:corner-c11-c13}  
\end{figure*}

\begin{figure*}
  \centering
    \includegraphics[width = 0.49\linewidth]{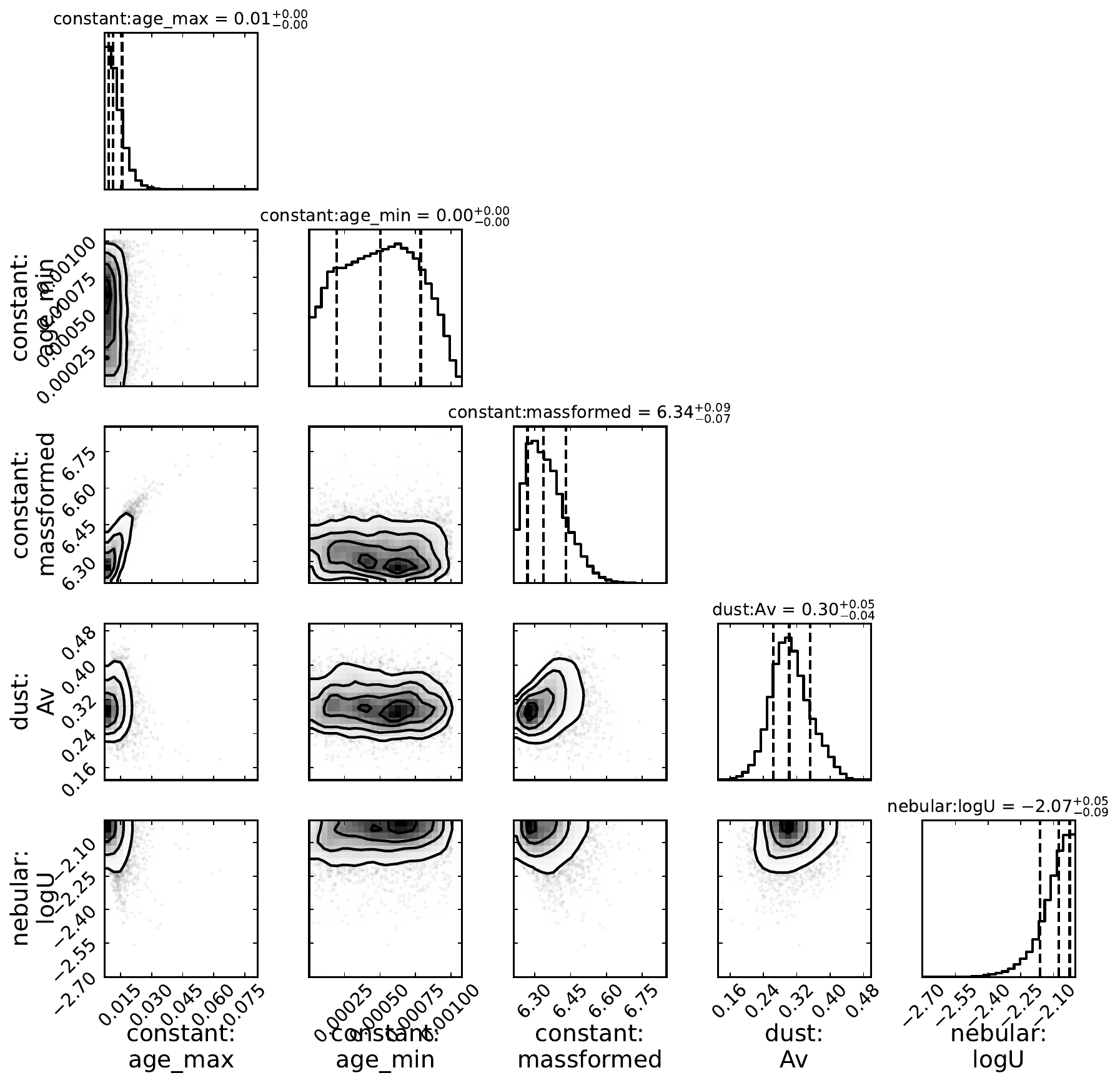}
       \includegraphics[width = 0.49\linewidth]{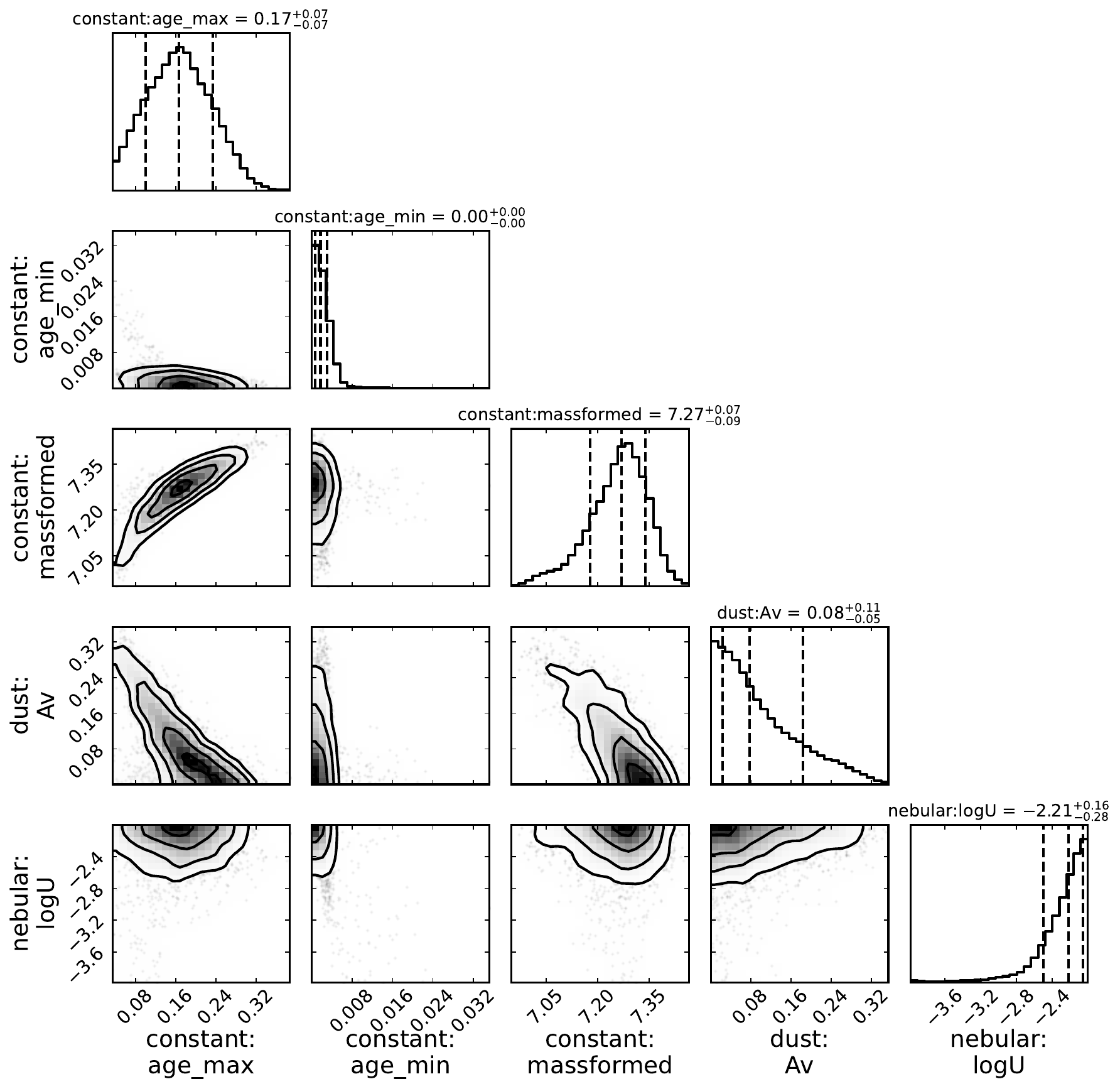}\\
         \includegraphics[width = 0.49\linewidth]{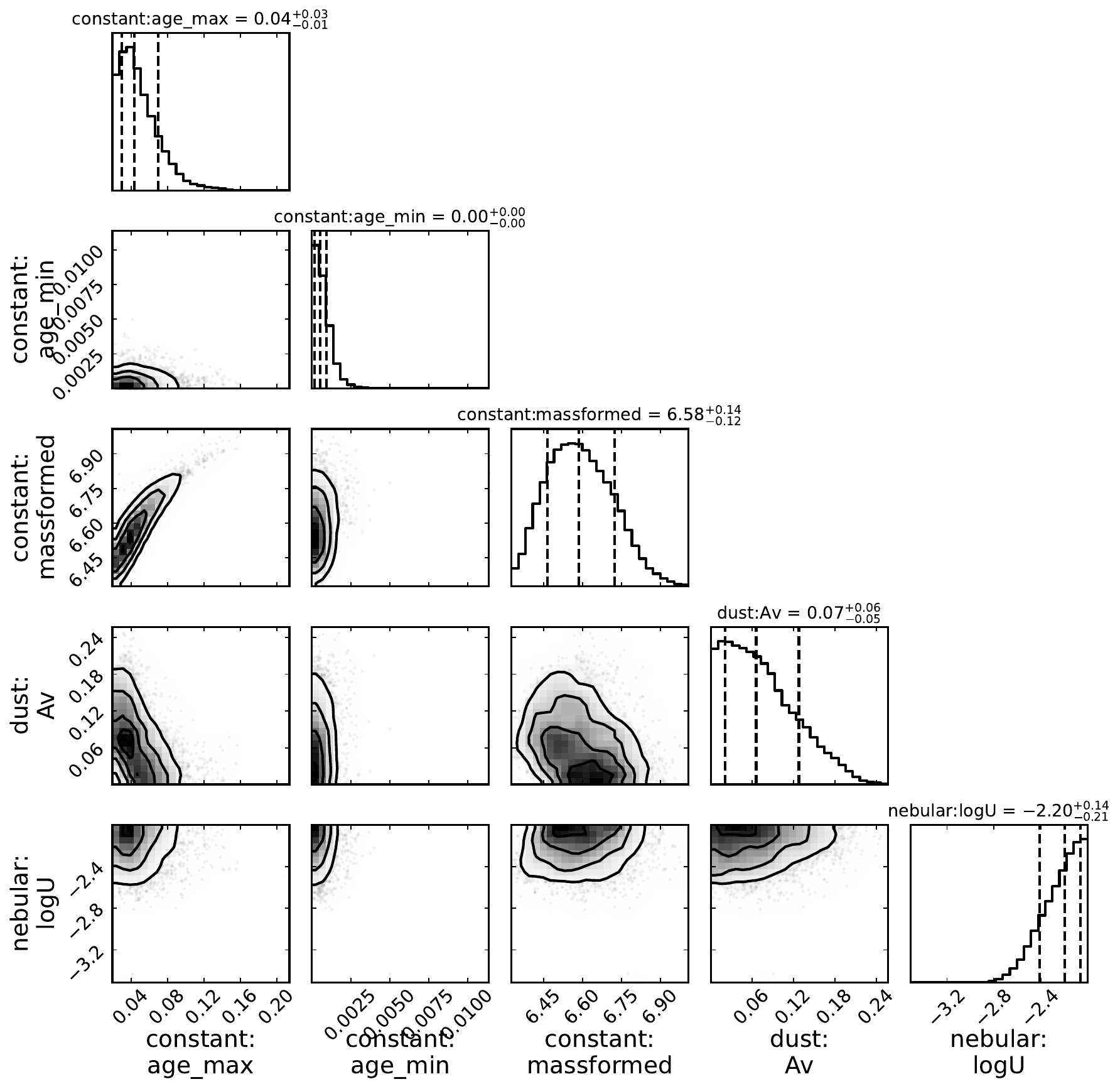}
       \includegraphics[width = 0.49\linewidth]{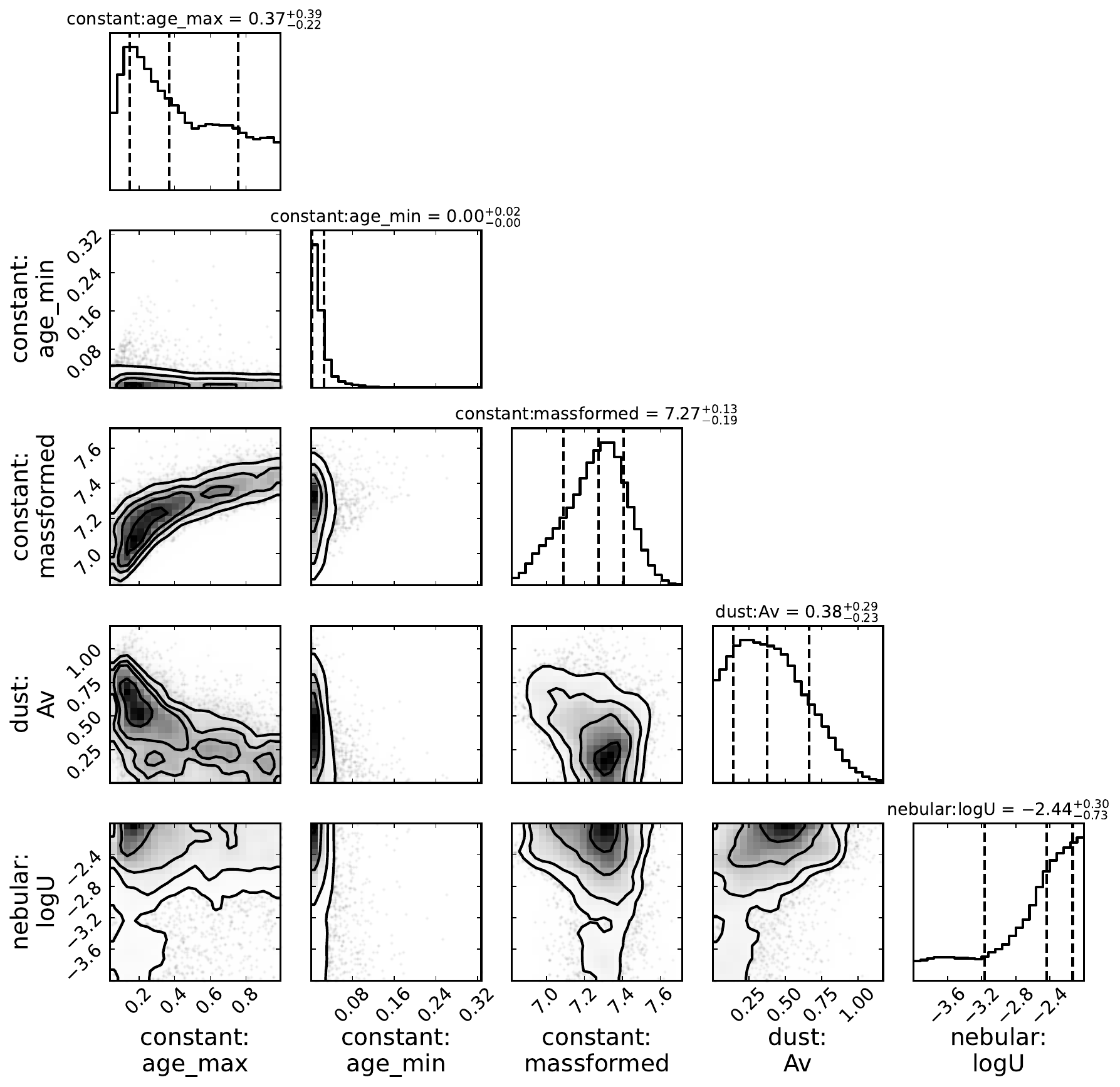}\\
    \caption{Corner plots derived from {\sc BAGPIPES} for CRISTAL-15 (top row) and CRISTAL-17 (bottom row) for the SED fitting of individual pixels.  A CSFH was assumed.
    The relevant pixels are highlighted in Fig.~\ref{fig:c11} and~\ref{fig:c17}, with the left column corresponding to the red highlighted pixels and the right column corresponding to the orange highlighted pixels.
    }
    \label{fig:corner-c15-c17}  
\end{figure*}

\section*{Affiliations}
\noindent
{\it
$^{1}$Jodrell Bank Centre for Astrophysics, University of Manchester, Oxford Road, Manchester, UK \\
$^{2}$Institute of Cosmology and Gravitation, University of Portsmouth, Burnaby Road, Portsmouth, PO1 3FX, UK\\
$^{3}$Sub-department of Astrophysics, University of Oxford, Denys Wilkinson Building, Keble Road, Oxford, OX1 3RH, UK\\
$^{4}$ Department of Physics, The University of Tokyo, 7-3-1 Hongo, Bunkyo, Tokyo 113-0033, Japan\\
$^{5}$Instituto de Estudios Astrof\'isicos, Facultad de Ingenier\'ia y Ciencias, Universidad Diego Portales, Av. Ej\'ercito Libertador 441, Santiago, Chile\\
$^{6}$Department of Physics and Astronomy and George P. and Cynthia Woods Mitchell Institute for Fundamental Physics and Astronomy, Texas A\&M University, \\4242 TAMU, College Station, TX 77843-4242, US\\
$^{7}$ Universidad Autonoma de Madrid, Ciudad Universitaria de Cantoblanco, E-28049 Madrid, Spain\\
$^{8}$  CIAFF, Facultad de Ciencias, Universidad Autonoma de Madrid, E-28049 Madrid, Spain\\
$^{9}$International Centre for Radio Astronomy Research, University of Western Australia, 35 Stirling Hwy, Crawley 26WA 6009, Australia\\
$^{10}$ARC Centre of Excellence for All Sky Astrophysics in 3 Dimensions (ASTRO 3D), Australia \\
$^{11}$Sterrenkundig Observatorium, Ghent University, Krijgslaan 281 - S9, B-9000 Gent, Belgium\\
$^{12}$Institute for Astronomy, University of Edinburgh, Royal Observatory, Edinburgh, EH9 3HJ, UK\\
$^{13}$Scuola Normale Superiore, Piazza dei Cavalieri 7, 56126 Pisa, Italy\\
$^{14}$Space Telescope Science Institute, 3700 San Martin Drive, Baltimore, MD 21218, USA\\
$^{15}$Departamento de Astronom\'ia, Universidad de Concepci\'on, Barrio Universitario, Concepci\'on, Chile\\
$^{16}$Department of Astronomy, School of Science, SOKENDAI (The Graduate University for Advanced Studies), 2-21-1 Osawa, Mitaka, Tokyo 181-8588, Japan\\
$^{17}$National Astronomical Observatory of Japan, 2-21-1 Osawa, Mitaka, Tokyo 181-8588, Japan\\
$^{18}$International Space Centre (ISC), The University of Western Australia, M468, 35 Stirling Highway, Crawley, WA 6009, Australia \\
$^{19}$Waseda Research Institute for Science and Engineering, Faculty of Science and Engineering, Waseda University, 3-4-1 Okubo, Shinjuku, Tokyo 169-8555, Japan\\
$^{20}$Centro de Astrobiolog\'{\i}a (CAB), CSIC-INTA, Ctra. de Ajalvir km 4, Torrej\'on de Ardoz, E-28850, Madrid, Spain \\
$^{21}$Dept. Fisica Teorica y del Cosmos, Universidad de Granada, Spain\\
$^{22}$  Kavli Institute for the Physics and Mathematics of the Universe (WPI), UT Institute for Advanced Study, The University of Tokyo, Kashiwa, Chiba 277-8583, Japan\\
$^{23}$ Research Center for the Early Universe, School of Science, The University of Tokyo, 7-3-1 Hongo, Bunkyo, Tokyo 113-0033, Japan

}


\bsp	
\label{lastpage}
\end{document}